\documentclass[12pt,usenames,dvipsnames,a4paper]{article}
\pdfoutput=1
\usepackage{soul}

\usepackage[utf8]{inputenc}
\usepackage[T1]{fontenc}

\usepackage{changepage}
\usepackage{amsmath,amsfonts,amssymb}
\usepackage{epsf,amsmath,bbold,amsfonts,stmaryrd}
\usepackage{mathrsfs}
\usepackage{appendix}
\usepackage{caption}
\usepackage{color}
\usepackage{datetime}
\usepackage{float}
\usepackage{graphicx}
\usepackage[colorlinks]{hyperref}
\hypersetup{pageanchor=false,citecolor=red,urlcolor=red}
\usepackage{indentfirst}
\usepackage[numbers,square,comma,sort&compress,merge]{natbib} 
\usepackage{subfig}
\numberwithin{equation}{section}
\usepackage{mathtools}
\usepackage[colorinlistoftodos]{todonotes}
\usepackage{ytableau}
\usepackage{tabu}
\usepackage{esvect}

\usepackage{calc}

\allowdisplaybreaks

\newcommand\bw{\begin{widetext}}
\newcommand\ew{\end{widetext}}

 \def\be{\begin{equation}}
\def\ee{\end{equation}}
 \def\ba{\begin{align}}
\def\ea{\end{align}}
\def\bea{\begin{eqnarray}}
\def\eea{\end{eqnarray}}

\def\br{\begin{flushright}}
\def\er{\end{flushright}}

\begin{document}

\begin{titlepage}

\vspace*{-2.5cm}
\begin{adjustwidth}{}{-.45cm}
\br
\er
\end{adjustwidth}

\vspace*{1.5cm}
\begin{adjustwidth}{-1.3cm}{-.7cm}

\begin{center}
    \bf \Large{Sterile Neutrino Dark Matter, \\Matter-Antimatter Separation, \\ and the QCD Phase Transition}
\end{center}
\end{adjustwidth}

\begin{center}
\textsc{Mikhail Shaposhnikov$^\star$ and ~Alexei Yu Smirnov\,$^\dagger$}
\end{center}

\begin{center}
\it {$^\star$Institute of Physics \\
\'Ecole Polytechnique F\'ed\'erale de Lausanne (EPFL) \\ 
CH-1015 Lausanne, Switzerland

\vspace{.4cm}

$^\dagger$Max-Planck-Institut f\"ur Kernphysik, Saupfercheckweg 1, \\
69117 Heidelberg, Germany
}
\end{center}

\begin{center}
\small
\texttt{\small $^\star$mikhail.shaposhnikov@epfl.ch} 
\\
\texttt{\small $^\dagger$smirnov@mpi-hd.mpg.de} 
\end{center}

\begin{abstract}

The Universe may contain sufficiently small size matter-antimatter domains at temperatures of a few hundred MeV, without violating the success of Big Bang Nucleosynthesis. We demonstrate that this possibility enhances the keV scale sterile neutrino production and may lead to its abundance consistent with the observable energy density of dark matter (DM).  We suggest that the separation of matter and antimatter,  creating temporarily macroscopic domains occupied by hadronic matter and quark-gluon plasma with an excess of baryons over anti-baryons and vice versa largely exceeding the average baryon and lepton asymmetries of the Universe, may appear because of the first-order QCD phase transition. Although the lattice studies provided a piece of evidence in favour of a  smooth crossover between the hadronic and quark-gluon phases at high temperatures and zero chemical potential for baryonic number, we argue that these simulations might not rule out relatively weekly first-order phase transition.  We discuss several scenarios of matter-antimatter separation at the QCD phase transition and the production of DM sterile neutrinos in each of them. One of the possibilities requires the presence of lepton asymmetry of the Universe, which can be smaller than that needed for the DM correct abundance in the homogeneous case.
\end{abstract}

\end{titlepage}

\tableofcontents

\section{Introduction}
\label{intro}

The motivation for the sterile neutrino Dark Matter (DM)  candidate \cite{Dodelson:1993je,Shi:1998km}  stems from discovery of neutrino oscillations \cite{Cleveland:1998nv,SAGE:1999nng,SAGE:2002fps,GALLEX:1998kcz,GNO:2000avz,Super-Kamiokande:2001ljr,SNO:2001kpb,SNO:2002tuh,KamLAND:2002uet,KamLAND:2004mhv,Super-Kamiokande:1998kpq,Super-Kamiokande:2004orf,Soudan-2:1999jbo,MACRO:2001fie,K2K:2004iot}.  The simplest and most natural way to get non-zero neutrino masses is to add to the Standard Model several (at least two) right-handed neutrinos  (or, what is the same, Majorana fermions or Heavy Neutral Leptons-HNLs) \cite{Minkowski:1977sc,Yanagida:1979as,Gell-Mann:1979vob,Glashow:1979nm,Mohapatra:1979ia}.  These particles have Yukawa couplings with the Higgs boson and leptonic doublets,  which generate the active neutrino masses via the see-saw mechanism.  If the number of HNLs is three (the same as the number of fermionic generations in the Standard Model)  this model, known as the minimal type I see-saw model or the $\nu$MSM  (neutrino minimal Standard Model) \cite{Asaka:2005an,Asaka:2005pn},  is capable of explaining simultaneously the Dark Matter (in terms of the lightest HNL - $N_{1}$ with the mass in the keV range), and also neutrino masses and baryon asymmetry of the Universe coming about because of heavier HNLs $N_{2,3}$,  addressing thus all experimental drawbacks of the SM.  The keV sterile neutrinos do not contribute to the masses of light neutrinos, which allows us to fix the absolute scale of active neutrino masses \cite{Asaka:2005an} within the  $\nu$MSM.

In the  $\nu$MSM, the DM sterile neutrinos are most effectively produced via mixing with active neutrinos in reactions with other particles of the SM,  such as $l^+l^-\to \nu N_1$, at temperatures of a few hundred MeV. \footnote{The DM sterile neutrino may also be created in the processes including physics beyond the $\nu$MSM. The proposals include the decays of extra scalar particles \cite{Shaposhnikov:2006xi,Kusenko:2006rh},  higher-dimensional operators \cite{Bezrukov:2011sz}, Einstein-Cartan 4-fermion  gravitational interaction \cite{Shaposhnikov:2020aen},  left-right symmetric theories \cite{Dror:2020jzy} etc.} In the case of small lepton asymmetries of the Universe (the Dodelson-Widrow mechanism \cite{Dodelson:1993je}),  the required mixing angle is in contradiction with X-ray\footnote{ $N_1$ is unstable and can decay as $N_1\to \gamma\nu$, producing a narrow X-ray line that can be seen by X-ray telescopes \cite{Abazajian:2001vt,Abazajian:2001nj,Dolgov:2000ew}.  Evidence for such a line at $3.5$ keV which would correspond to decays of  $7$ keV DM sterile neutrino was reported in \cite{Bulbul:2014sua,Boyarsky:2014jta}.  It remains to be seen if this line indeed corresponds to the radiative decay of DM particles.}  and structure formation constraints\footnote{The free streaming length of $N_1$ admitted by the X-ray constraints is too large and contradicts to the cosmological  Lyman-$\alpha$  forest data, for a review see \cite{Boyarsky:2018tvu}.}.  

If the lepton asymmetry is much larger than the baryon asymmetry \cite{Laine:2008pg,Ghiglieri:2015jua}, and furthermore 
\be
\label{defLcrit}
\Delta_L \gtrsim L_{\rm crit}  \equiv 6.6\times 10^{-5}\,,
\ee
then the production of the sterile neutrinos is resonantly enhanced according to the Shi-Fuller mechanism \cite{Shi:1998km}  and all the constraints mentioned above can be satisfied (for a review see \cite{Boyarsky:2018tvu}).  We define $\Delta_L=L/s$, where $L$ is the density of the total lepton number and $s$ is the entropy density. These quantities are taken at a temperature $4$ GeV, and it is assumed that the lepton numbers of different generations can only be changed due to the mixing with DM sterile neutrino (or, in other words, that the processes with $N_{2,3}$ of the $\nu$MSM are irrelevant). The number $L_{\rm crit}$ depends on the flavour composition of the lepton asymmetry, on the sterile neutrino mixing angle $\theta$, and on the mass of the DM sterile neutrino \cite{Laine:2008pg,Ghiglieri:2015jua}. The specific value of $L_{\rm crit}$ given above in (\ref{defLcrit}) corresponds to $L_e=L_\mu=L_\tau=L/3$,  $M_1=7$ keV and $\theta^2=5\times10^{-11}$  \cite{Ghiglieri:2015jua}. 

 The lepton asymmetries of this magnitude can indeed be produced in the $\nu$MSM,  albeit in a fine-tuned domain of the parameter space leading to strong degeneracy between heavier HNLs $N_2$ and $N_3$ \cite{Shaposhnikov:2008pf,Canetti:2012kh,Ghiglieri:2020ulj}. 

To the best of our knowledge, in all approaches to sterile neutrino production, the Universe was taken to be homogeneous and isotropic at the relevant temperatures of the order of $200$ MeV. However, this is a mere assumption. The success of the Big Bang Nucleosynthesis (BBN) indicates that this was very likely to be the case at smaller temperatures $T\sim 1$ MeV, but the inhomogeneities at $T\sim 200$ MeV with a size as large as few meters are admitted \cite{Campbell:1990ak}, as they dissipate before the BBN starts through the combined action of neutrino inflation and neutron diffusion \cite{Applegate:1987hm,Jedamzik:1993dc,Jedamzik:1994ux,Kurki-Suonio:1996wfr}. Moreover, even the existence of matter-antimatter domains with a baryon-to-entropy ratio of the order of one (or minus one) is allowed, provided their size is smaller than the neutron diffusion length at the BBN epoch. 

The presence of these domains may change considerably the sterile neutrino DM abundance and their momentum distribution, important for structure formation. Indeed, even though the average baryon asymmetry of the Universe is small, it can be large locally, leading to the resonant production of sterile neutrinos. Furthermore, the resonance will occur for the left chirality of sterile neutrinos in one type of domain, and with right chirality in another, producing the chiral symmetric DM, contrary to the resonance production in the homogeneous situation. 

How the matter-antimatter domains can appear in the Universe? One of the mechanisms suggested in the literature a while ago is associated with a possible existence of stochastic hypermagnetic fields at temperatures above the sphaleron freeze-out  \cite{Giovannini:1997gp} (see also \cite{Kamada:2016eeb,Kamada:2016cnb}). The electroweak anomaly converts the hypermagnetic fields into baryons and thus leads to an inhomogeneous Universe with matter-anti-matter domains. Yet another possibility to have matter-antimatter domains appears in theories with two sources of CP-violation - spontaneous and intrinsic \cite{Kuzmin:1981ip}. Also, the matter-antimatter domains may appear in inhomogeneous baryogenesis, described in \cite{Dolgov:1992pu,Dolgov:2008wu}.

Our current work gives yet another example, associated with the first-order QCD phase transition. We will show that in the scenario with matter-antimatter separation at the first-order QCD phase transition, the sterile neutrino DM production may be enhanced considerably. The enhancement can be efficient even if the asymmetry is small $\Delta_L < L_{\rm crit}$,   or even absent, $\Delta_L = 0$ if the Omnes type picture of the separation (discussed in section \ref{omnes}) is correct. Though we concentrate on these specific mechanisms of matter-antimatter separation at the QCD epoch, our findings about sterile neutrino DM generation are universal and applicable also to other possibilities, mentioned above.  In most numerical estimates, we take $M_1=7$ keV and $\theta^2=5\times10^{-11}$, but the equations we present are valid for the other choice of parameters as well. 

To sum up, the aim of the present paper is threefold. First, we will argue that the conclusions about the absence of the QCD phase transition (for a recent review see \cite{Aarts:2023vsf})  might be still premature.  The lattice simulations are carried out necessarily with finite lattice spacing and in the finite volume. The latter means that it is difficult to rule out the existence of a weak first-order QCD phase transition, with a relatively small latent heat $l\lesssim \Lambda ^4$ and interphase tension $\sigma\lesssim \Lambda ^3$  between the coexisting QGP (quark-gluon plasma) and hadronic phases. With the use of some very rude modelling of the finite volume effects, we make an estimate of these parameters,  which might still be allowed by the lattice simulations.
 
Second, assuming that the QCD phase transition indeed takes place, we argue that it may lead to temporary matter-antimatter separation,  creating macroscopic domains (and ``anti-domains'') with an excess of baryons over anti-baryons (and vice versa) substantially exceeding the average baryon and lepton asymmetries of the Universe.

Finally, we uncover a new possible consequence of the first-order QCD phase transition, associated with the production of sterile neutrino DM.   We will see that the inhomogeneities produced at this transition may facilitate the otherwise suppressed creation of sterile neutrinos.  The droplets of QGP rich in baryon number formed at the first order QCD phase transition play a crucial role in the mechanism of sterile neutrino DM production we propose.  We will see that the creation of sterile neutrinos via the conversion (oscillations) of ordinary neutrino $\nu$ into $N_1$  inside the droplets may lead to the DM abundance consistent with observations.  In addition,  after the phase transition, the same type of production processes occur in the hadronic phase inside the lumps with the excess of the baryon number,  slowly spreading when the Universe expands.  The enhancement of sterile neutrino production is similar to the MSW effect in neutrino oscillations in the medium \cite{Mikheev:1986wj}. 

The paper is organised as follows. In Section \ref{phase},  we consider possible scenarios of matter-antimatter separation at the QCD phase transition. We start with a discussion of the finite volume effects which might obscure the nature of the first-order phase transitions in lattice simulations.  In subsection \ref{QCDPT}, we get a glimpse of its possible parameters a get a rough estimate of the average distance between nucleating bubble centres and the probability distribution of the droplets of quark-gluon plasma as a function of their sizes  (using the previous studies of the QCD phase transition  \cite{Kajantie:1986hq,Ignatius:1994fr}).  Subsection \ref{LA} discusses the mechanism(s) of matter-antimatter separation in the first-order QCD phase transition. In Section \ref{sterile} we investigate the sterile neutrino production in the scenarios with matter-antimatter separation.  We study first the resonance $N$  production in a homogeneous medium and apply it to Omnes phase transition and QGP droplets of big size. The case of small droplets is considered in subsection \ref{smalldrop}. In the last Section, we summarise our results. 

\section{QCD phase transition and matter-antimatter  separation} \label{phase}

There is a common belief that the evolution of the Universe in the framework of the Standard Model of elementary particles is smooth. Potentially, during the Universe cooling after the Big Bang, one could encounter two phase transitions. The first one is the electroweak phase transition (EWPT) at $T\simeq 160$ GeV passing from the symmetric to the Higgs phase. The second one is the QCD phase transition at $T\simeq 160$ MeV from the quark-gluon plasma to the gas of hadronic states. In both cases, there is no true gauge-invariant order parameter that can distinguish relevant phases, meaning that in general the phase transitions are either absent (smooth cross-over), or are of the first order. A specific choice of parameters of the theory may lead to a second-order phase transition (for example, in the EW case this would happen if the Higgs boson mass were $M_H \approx 73$ GeV \cite{Kajantie:1996mn,Kajantie:1996qd}). In both cases, the nature of the phase transitions cannot be resolved by the use of perturbation theory, and one has to use non-perturbative methods such as lattice simulations. 

The precision study of the EWPT can be done by a combination of perturbative and non-perturbative computations. In the first step, one constructs an effective {\em three-dimensional purely bosonic theory}  containing the gauge bosons of the SU(2)$\times$U(1) group and the Higgs doublet \cite{Kajantie:1995dw}.  This effective theory is a lot simpler than the original four-dimensional Standard Model: there are no fermions and no strong SU(3) interactions; moreover, the effective theory is super-renormalisable. All these allow for very accurate lattice simulations. This program led to the result that for the experimentally measured value of the Higgs mass $M_H \approx 125$ GeV, the transition from the symmetric to the Higgs phase in the SM is a smooth crossover \cite{Kajantie:1996mn,Kajantie:1996qd,DOnofrio:2015gop} (it would be a first-order phase transition if the Higgs mass were below $73$ GeV)\footnote{The four-dimensional lattice simulations of the bosonic sector of the EW theory \cite{Csikor:1998eu,Aoki:1999fi} confirmed this conclusion.}. 

The study of the QCD phase transition is way more complicated. At the temperatures of the order of the confinement scale $\Lambda \sim 200$ MeV, where the phase transition (if any) is expected to take place, the system is strongly coupled, and no three-dimensional effective bosonic description is possible. In addition, there are technical challenges of putting on the lattice nearly massless quarks, such as $u$ or $d$, and eventually $s$. Still, the lattice studies \cite{Aoki:2006we,Bazavov:2009zn,Borsanyi:2010cj,Bhattacharya:2014ara} reported a piece of evidence for the smooth cross-over, cutting off the discussions of cosmological applications of the first-order QCD phase transition, such as black hole formation, gravitational wave generation,  non-homogeneous nucleosynthesis, amplification of cosmological density perturbations, to list a few. 

\subsection{QCD phase transition}
\label{QCDPT}

\begin{figure}[h]
   \centering
    \includegraphics[width=0.46\textwidth]{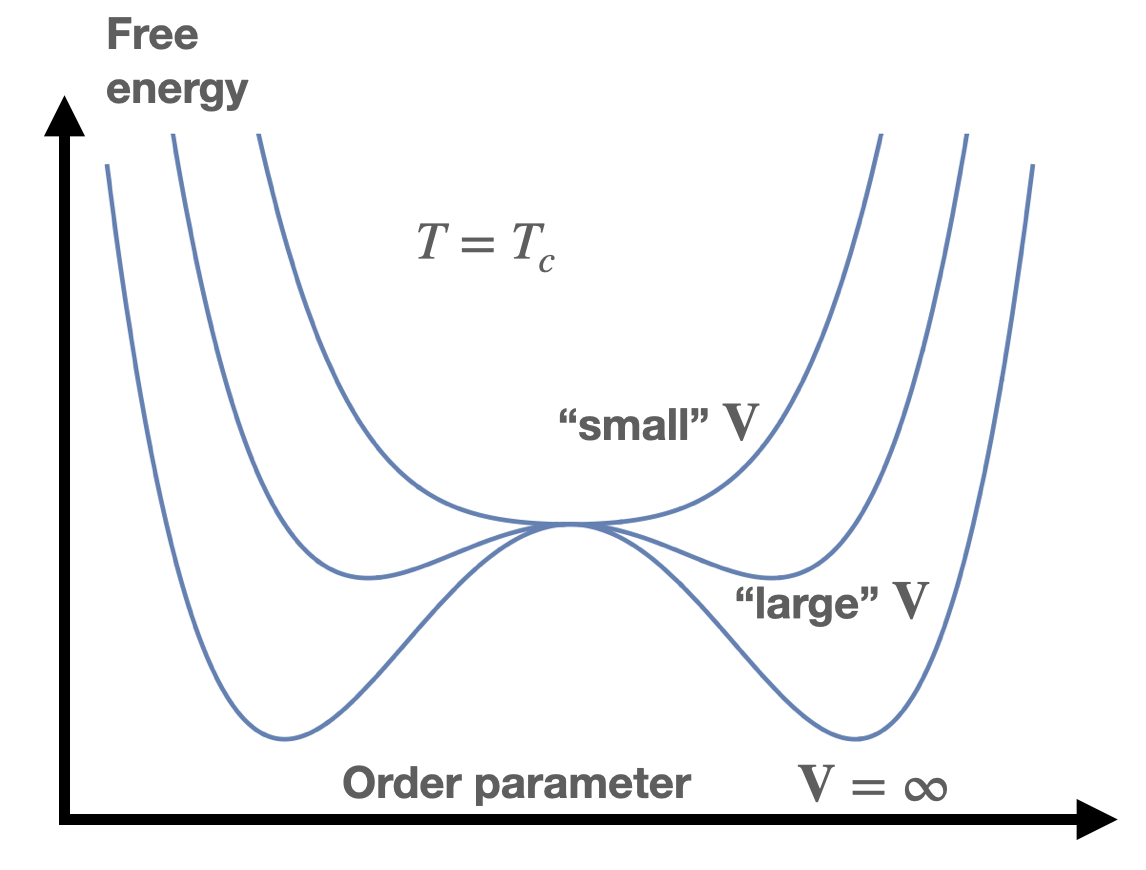}
    \caption{The effective potential at the critical temperature for a first-order phase transition. It has two separated minima at infinite or ``large'' volumes of the system. The double-phase structure disappears at small volumes.}
    \label{fig:effpot}
\end{figure}

It is textbook knowledge that free energy is an analytic function of temperature if the volume $V$ of the system is finite.  In other words, there are no phase transitions when $V<\infty$. Still, the lattice studies do allow us to trace the simultaneous existence of different minima of the effective potential for an order parameter, provided they exist at finite volume.  It may happen, however, that the finite volume and lattice spacing effects change the form of the effective potential substantially removing the double-phase structure, see Fig. \ref{fig:effpot}. In this case, if made at ``small'' volumes and ``large'' lattice spacings, the lattice simulations will report the smooth cross-over rather than the first-order phase transition.

Due to strong coupling, what happens in QCD with the change of the volume and lattice spacing, is a complicated question. A hint that the QCD phase transition in the infinite volume may be of the first order comes from phenomenological studies based on the MIT bag model for hadronic states and its generalisations (see, e.g. \cite{Kajantie:1986hq,Ignatius:1994fr}). In what follows, we take a relatively simple theory (having nothing to do with the real QCD) which can be treated by perturbation theory, in which the volume and lattice spacing dependence can be studied reliably. That will give us a rough idea of the allowed characteristics of the phase transitions (latent heat and the interphase tension) when the lattice simulations show the cross-over.

Let us consider the SU(2) gauge theory with the scalar doublet at finite temperature $T$. It depends on 3 parameters - the gauge coupling $g$, the scalar self-coupling $\lambda$, and the mass of the scalar particle $m^2$. For $g$ and $\lambda$ sufficiently small the phase transitions can be studied in the effective 3-dimensional Euclidean theory, by integrating out the Matsubara modes with masses $\sim T$ and $\sim g T$ \cite{Kajantie:1995dw}. The effective Lagrangian has the form
\be
  L_{\rm eff}={\scriptstyle{1\over4}} F_{\mu\nu}^aF_{\mu\nu}^a +
  (D_\mu\phi)^\dagger (D_\mu\phi) +\mu_3^2 \phi^\dagger\phi + \lambda_3
  (\phi^\dagger\phi)^2, 
\label{3daction}
\ee
where $F_{\mu\nu}$ is the SU(2) field strength, $D_\mu$ is the covariant derivative, the 3D couplings are $g_3^2=g^2 T,~\lambda_3=\lambda T$ and the 3D mass parameter is $\mu_3^2\equiv \gamma(T^2-T_0^2)=m^2+\left(\frac{3}{16}g^2+\frac{\lambda}{2}\right)T^2$ (we omit the log-type corrections irrelevant for the present discussion). The continuum one-loop effective potential for $\varphi^2=2\phi^\dagger\phi$ is
\be
V_{\rm eff}= \frac{1}{2} \mu_3^2 \varphi^2 -\frac{T}{16\pi}g^3\varphi^3+\frac{\lambda}{4} \varphi^4~,
\label{Veff}
\ee
where the $\varphi^3$ term comes from the gauge loops (we assumed that $\lambda \ll g^2$). This theory has a first-order phase transition at $T_c^2=T_0^2/(1-\lambda \kappa^2/(2\gamma))$, the jump of the order parameter $\Delta\varphi=\kappa T_c$ (to be compared with the zero-temperature vacuum expectation value  $\langle \varphi \rangle = T_0\sqrt{\gamma/\lambda}$), the masses of the scalar and vector excitations $m_s= \kappa T_c \sqrt{\lambda/2},~~m_v= \kappa T_c g/2$  respectively, the latent heat $l=\gamma\kappa^2 T_c^2T_0^2$, and the interphase tension $\sigma=\Delta\varphi^3\sqrt{\lambda}/(6\sqrt{2})$, where $\kappa=g^3/(8\pi\lambda)$. The two phases can coexist for temperatures $T$ within the interval  $T_-<T<T_+$, where $T_+^2=T_0^2/(1-9\lambda \kappa^2/(16\gamma))$ and $T_-=T_0$.

\begin{figure}[tb]
    \centering
    \includegraphics[width=0.46\textwidth]{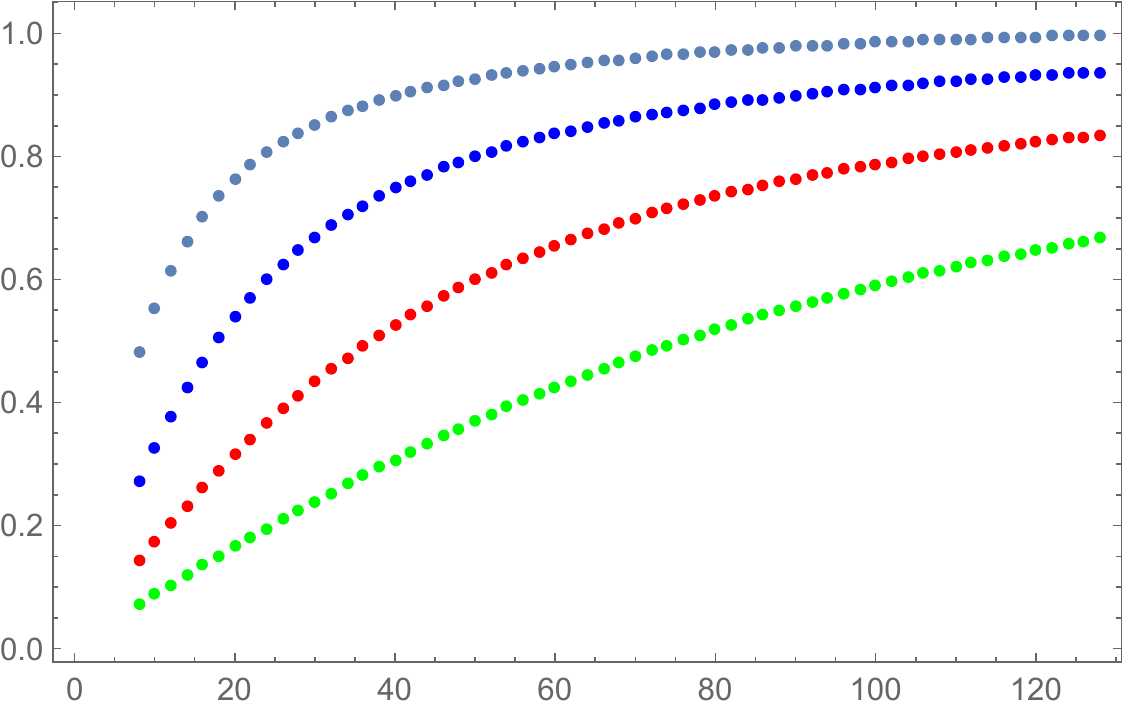}
    \caption{Verical axis: the ratio of the cubic term on the lattice with size $N^3$ to its continuum value for  $m_v a=0.24$ (black), $m_v a=0.12$ (blue), $m_v a=0.06$ (red)  and  $m_v a=0.03$ (green) dots. Horizontal axis: the lattice size $N$.}
    \label{fig:Ndep}
\end{figure}

As was shown in \cite{Kajantie:1993ag}, this system develops a first-order phase transition on a  $N^3$ cubic lattice with periodic boundary conditions only if the equation
\be
\lambda={3g^4\over128}aT{1\over N^3} \sum_{n_i=0}^{N-1}
{1\over d[d+(ga\phi/4)^2]} 
\label{cond}
\ee
has a solution for $\phi$. Here $a$ is the lattice spacing, and
\be
d =\sin^2(\pi n_1/N)+\sin^2(\pi n_2/N)+\sin^2(\pi n_3/N)\, .
\ee
Using the value of the vector boson mass and the lattice size $L= N a$ the solution appears at lattice sizes exceeding
\be
m_v L > 4.993~.
\label{constr}
\ee
This result may seem surprising: the finite size effects are expected to be suppressed exponentially, as $\exp{(-L/l_{\rm corr})}$, where $l_{corr}$ is the largest static correlation length in the system. Taking it to be $l_{\rm corr} =1/m_v$, one gets from eq. (\ref{constr})  an estimate of the finite volume correction at the level of 1\%, $\exp{(-L/l_{\rm corr})} \simeq 7\times10^{-3}$, whereas in reality the use of the finite volume and lattice spacing even changes the qualitative behaviour of the system (\ref{3daction}).  In Fig. \ref{fig:Ndep} we present the ratio of the lattice to continuum values of the cubic term for $m_v a=0.03,~0.06$,  $0.12$ and $0.24$ (where $a$ is the lattice spacing, and these particular values correspond to the variety of lattice spacing used in the QCD simulations if $m_v$ is identified with the pion mass). Even for huge lattices with $N=128$, the error in this ratio may be about tens per cent.  

The lattice simulations of  \cite{Aoki:2006we,Bazavov:2009zn,Borsanyi:2010cj}  were done at $L T_c = 3,4,~{\rm and}~5$. Taking $T_c\simeq 155$ MeV, and $1/m_\pi\simeq 1/(139~\rm{MeV})$ for the correlation  length\footnote{The static pion screening mass is known to be somewhat larger at non-zero temperatures \cite{Brandt:2015sxa}, but this does not change qualitative conclusions.}, we get that for these simulations the product $m_\pi L$ was $2.7,~3.6$ and $4.5$ respectively. The analysis of \cite{Bhattacharya:2014ara} was done for volumes $(4~{\rm Fermi})^3$ and $(11~{\rm Fermi})^3$, giving  $m_\pi L= 2.8$ and $7.8$. 

Of course, the physics of the QCD phase transition has nothing to do with the example we used, but it is alarming to see that the criterion (\ref{constr}) was not satisfied for the majority of simulations and barely exceeded it just for a few of them. We tend to conclude from here that the lattice simulations with considerably larger volumes might be needed to elucidate the nature of the QCD phase transition. They should include also the analysis of static correlation length in channels with different quantum numbers, to track the possible finite size effects.

How large the necessary lattice sizes should be? A possible (naive) estimate comes from the requirement to track the presence of baryons in the lattice volume. If one uses the zero-temperature proton mass at $T_c$, one finds that the lattice with the size $L T_c = 4$ and $N=64$ usually used in the simulations accommodates just $0.16$ spin states of the proton. To have a reasonable presence of other hadrons, such as, say, two $\Lambda$'s inside the lattice volume to account for strangeness,  one would need $L T_c \sim 12$, and thus the spacial lattice sizes as large as $128^3$. We also remark here that the lattice size in the temporal temperature direction is usually taken to be $N_t=8$, which can only accommodate just two non-trivial Matsubara harmonics along the imaginary ``time'' direction. 

From now on we take the liberty to assume the QCD phase transition is of the first order and study how it may develop in the early Universe.

\subsection{Cosmic separation of phases}
\label{separation}

Let us assume for the time being that there are just two phases - hadronic and QGP that may coexist in some interval of temperatures  $T_-<T<T_+$. A  more intricate and very speculative possibility will be discussed in section \ref{omnes}. 

We start from a short overview of the previous works on this problem, relevant to us. We assume in what follows that before the QCD phase transition the Universe is homogeneous and contains a tiny baryon asymmetry, $\Delta_B \equiv  B/s\simeq 9\times 10^{-11}$ ($ B$ and $s$ are the densities of the baryon number and entropy respectively) and, perhaps, the comparable lepton asymmetry. As was argued by Witten \cite{Witten:1984rs} the first-order QCD phase transition may lead to cosmic separation of hadronic and QGP phases. When the universe supercools somewhat below the critical temperature $T_c$ (we take it to be  $T_c\simeq 160$ MeV for numerical estimates) the bubbles of new, hadronic phase nucleate.  The shock waves originating from the bubble expansion reheat quickly the quark phase up to the critical temperature, and the universe stays at a constant temperature $T_c$ during the sizeable fraction of the Hubble time until the end of the transition. The hadron bubbles grow slowly and roughly with the Hubble rate $H$. They start to percolate at the moment $t_0$ when their fraction reaches $\sim 50\%$ of the space. After that, the situation is reversed - the hadronic phase is dominating, and the droplets with the quark-gluon plasma inside shrink slowly. 

This process may lead to the separation of the net baryon number \cite{Witten:1984rs}:  the droplets with the plasma are much richer in baryon number than the baryonic phase, because quarks are lighter than hadrons and the transport of the baryon number over the phase boundary is suppressed \cite{Witten:1984rs,Jedamzik:1994tmi}. It was conjectured in \cite{Witten:1984rs} that the baryon asymmetry inside the droplets can be of the order of one at the end of the phase transition.  

The dynamics of the bubble nucleation in QCD phase transition in this case is mainly determined by the latent heat $l$ and the interphase tension $\sigma$. It was considered in detail in \cite{Kajantie:1986hq}, and we just give here the main formulas from this article. 

First, the Universe is supercooled in the QGP state till temperature $T^*$ ($T^*$ is defined as a moment when the shock fronts originated on the nucleated bubbles start to collide). The amount of supercooling is characterised by 
\be
x^2=\left(1-\frac{T^*}{T_c}\right)^2 \simeq A\left[4 \log\left(\frac{2M_0}{T_c}\left(\frac{2\pi v^3}{3A}\right)^{\frac{1}{4}}\right)\right]^{-1}~,
\label{tstar}
\ee
where $M_0\simeq M_P/(1.66 {g^*}^{\frac{1}{2}}),~M_P=1.22\times 10^{19}$ GeV, $g^* = g^*_w+g^*_{\rm QCD}$ is the number of effectively massless degrees of freedom ($g^*_{\rm QCD}$ corresponds to the QCD degrees of freedom, and $g^*_w$ counts the rest), $v^2\approx 1/3$ is the sound speed and $A$ is the combination of $l$ and $\sigma$,
\be
A=\frac{16\pi \sigma^3}{3l^2 T_c}~.
\label{A}
\ee
At $T=T_c\simeq 160$ MeV  we take, following \cite{Laine:2006cp}, $g^*_w\simeq 14.25,~g^*_{\rm QCD}\simeq 10$, leading to $M_0\simeq 1.4\times 10^{18}$ GeV and to the horizon size of the Universe  $t_H=M_0/T^2\simeq 10^6$ cm.

The average distance between the nucleated bubbles of the hadronic phase is given by
\be
r_0\simeq t_H \frac{x^3}{A}~,
\ee
and the probability distribution of the bubbles on their size $r$ is roughly
\be
d {\cal P} = \exp\left(-\frac{r}{r_0}\right) \frac{dr}{r_0}.
\ee
After the bubbles start to percolate, the picture is inverted:  the hadronic phase becomes dominant, and the droplets of the QGP, having initially the average size $\sim r_0$ will be slowly shrinking and disappear eventually at the end of the phase transition, after the time of the order of the age of the Universe at the QCD phase transition. 

A crucial parameter in this picture is the distance between the nucleated bubbles $r_0$, estimated in \cite{Kajantie:1986hq} as $r_0 \simeq 100$ cm, four orders of magnitude smaller than the horizon size. We can also take our toy model (\ref{3daction}) and try to choose its parameters in such a way that it imitates the QCD mass scales, and has a first-order phase transition, but this phase transition is not seen in the lattice simulations. To this end, we can take $m_v \simeq m_\pi$ and  $T_c \simeq 160$ MeV.  Then, by introducing $\lambda = \zeta g^2$, one can see that all the relevant parameters, such as latent heat and bubble surface tension become the functions of $\zeta$. Varying $\zeta$ in a rather ad-hoc interval between $0.1$ and $1$, we get that $r_0$ changes from 20 cm to 1 cm respectively, not very far from an estimate of \cite{Kajantie:1986hq}.  It would be not unreasonable to assume that $r_0$ is somewhere between 1 cm and 1 meter. As we have already mentioned, this scale of inhomogeneities is too small to change the picture of the standard BBN.

If the baryon number is indeed confined inside the droplets of QGP,  the baryon number density in the droplets, $n_B^d(t)$, is given by  the solution of the equation
\be
\label{nBeq}
\frac{\partial n_B^d(t)}{\partial t} = -\frac{3\dot r}{r} n_B^d(t)~,
\ee
leading to the obvious solution
\be
\label{nBtime}
n_B^d(t) = n_B^d(t_0)\left(\frac{r(t_0)}{r(t)}\right)^3~,
\ee
where $r(t)$ is the time-dependent droplet size,  decreasing towards the end of the phase transition and $n_B^d(t_0) = \Delta_B \times s$  is the initial average baryon number density.

Having this basic picture in mind, we now extend it to the case when in addition to tiny baryon asymmetry the universe contains a sizable lepton asymmetry $\Delta_L\equiv L/s$, several orders of magnitude larger than the baryon asymmetry but still small enough to contradict different bounds coming, for instance, from Big Bang Nucleosynthesis (BBN) \cite{Rehm:1998nn}. The values as large as $\Delta_L\simeq 10^{-4}$ (i.e. million times more than the baryon asymmetry) can be generated, for example, in interactions of Heavy Neutral Leptons (HNLs) of the $\nu$MSM below the freeze out of sphaleron transitions, thus having no influence on baryon asymmetry of the Universe \cite{Shaposhnikov:2008pf,Canetti:2012vf,Canetti:2012kh,Ghiglieri:2020ulj,Eijima:2020shs}. If the lepton asymmetry of the Universe is close to its upper bound imposed by BBN, it can even change the standard cross-over picture of the QCD phase transition converting it to a first-order phase transition \cite{Wygas:2018otj,Schwarz:2009ii}. We are not going to consider this possibility in our work.

\subsection{Lepton Asymmetry and Matter-Antimatter separation}\label{LA}

For the following discussion, it is important to know the rates of the different weak reactions in the primordial plasma at the time of the QCD phase transition. The electron neutrino mean interaction rate coming from the scattering on leptons is given by  \cite{Notzold:1987ik}  
\be
\label{gammanu}
\Gamma_{\nu_e} = \frac{13}{9}\frac{7\pi}{24} G_F^2 T^4 \omega~,
\ee
where $G_F$ is a Fermi constant, and $\omega$ is the neutrino energy. A similar value is obtained for the muon neutrino.  For $\tau$ neutrino the rate is somewhat smaller due to the exponential suppression of $\tau$-lepton concentration. The presence of quarks in QGP and mesons and baryons in the hadronic phase changes these estimates (for discussion of hadronic uncertainties see \cite{Asaka:2006rw,Asaka:2006nq,Laine:2008pg,Ghiglieri:2015jua}) in an amount irrelevant to the present discussion.  For a typical neutrino energy, $\omega \sim 3T_c$ the neutrino mean free path (including their scattering on quarks) is $\lambda_\nu \sim 0.4$ cm, about six orders of magnitude smaller than the horizon size $\sim 10$ km at this time.

The rates of quark flavour non-conservation due to reactions of the type $\bar{u}d\to \mu^+\nu_\mu$, $\bar{u}s\to \mu^+\nu_\mu$ in QGP phase, and similar to the electron flavour,  is difficult to estimate reliably due to the strong coupling. It is expected that the number of effectively massless degrees of freedom in the QCD plasma at $T_c \sim 160$ MeV goes from   $g^*_{\rm free}=47.5$ (accounting for massless gluons and $u,d$ and $s$ quarks) to $g^*_{\rm int} \simeq 10$ \cite{Laine:2006cp}. As the interaction rate is proportional to the concentration of particles, it is reasonable to expect that the weak mean free path of $u$ and $d$ flavours is some factor $\sim 5$ larger than that of neutrinos, $\lambda_{u,d} \sim 2$ cm, and factor $\sim 100$ larger for $s$-quark, accounting for the Cabibbo angle suppression by $\sin^2\theta_c\simeq 0.05$,  $\lambda_s \sim 40$ cm. We expect to have similar estimates in the hadronic phase due to reactions of the type $\pi\pi\to \pi \nu e^+,~\pi K\to\pi\nu\mu^+$ and alike (the rates of 2-body decays of pions and kaons are suppressed either by chirality conservation or by the amount of the available phase space). 

The specific numbers for the mean free path given above are not that important in what follows. What is relevant for us is that the transitions between different hadronic flavours are well in thermal equilibrium.  In other words, an equilibrium plasma in the vicinity of the QCD phase transition is characterised by five chemical potentials $\mu_B,\mu_i$, and $\mu_Q$, corresponding to five conserved numbers: $B$ - baryon number, $L_{e,\mu,\tau}$ - 3 different lepton numbers, and electric charge $Q$. Imposing the electric neutrality of the QGP plasma, and neglecting for simplicity the masses of the light quarks $u,~d,~s$ and leptons $e,~\mu$ and any interactions between them one can easily find in linear order in asymmetries:
\be
\label{neutr}
\frac{\mu_B}{T}= \frac{3B}{T^3},~\frac{\mu_Q}{T} = 
\frac{3n_{L_e}+3n_{L_\mu}}{4T^3},~\frac{\mu_1}{T} 
= \frac{5n_{L_e}+n_{L_\mu}}{2T^3},~\frac{\mu_2}{T} 
= \frac{n_{L_e}+5n_{L_\mu}}{2T^3},~\frac{\mu_3}{T}= \frac{6n_{L_\tau}}{T^3}~,
\ee
where $n_i$ are the number densities of the corresponding conserved numbers in the obvious notations.  In this approximation, the asymmetries in the number of $d$ and $s$ quarks are the same. If the mass of $s$-quark is included, this degeneracy is broken. The analogous equations can be derived in the hadronic phase in the non-interacting gas approximation, with the use of baryons and mesons instead of quarks.

These relations show that in the presence of substantial lepton asymmetries $n_{L_i} \gg B$ the asymmetries in the individual quark flavours are of the order of $n_{L_i}$, rather than $B$. The importance of lepton asymmetries to the description of the QCD plasma was noted already in \cite{Zarembo:2000wj}. A similar picture arises in the hadronic phase, with excess of protons over antiprotons and mesons over anti-mesons of the order of lepton asymmetries. The physics of this phenomenon is obvious and is associated with the electric neutrality of the plasma: the equilibrium character of the weak interactions redistributes the leptonic asymmetry between neutrinos and electrically charged leptons, and the asymmetries in hadronic flavours are created to compensate for the charge imbalance.

We are coming now to our main observation. The non-zero and relatively large  (of the order of lepton asymmetry) chemical potential $\mu_Q$ breaks discrete symmetries C, CP,  and CPT that distinguish hadronic matter and antimatter, both in the QGP and hadronic phase.  In addition, the mass of the strange quark and masses of strange mesons and baryons are the order of the critical temperature or larger, which breaks the SU(3) flavour symmetry. It is plausible to think that these breakings are transmitted to the interaction of quarks and hadrons with the interphase boundary between the different phases\footnote{Of course, the C and CP-symmetries are also broken by the weak interaction due to Kobayashi-Maskawa CP-violating phase. We do not expect this to play any role in the matter-antimatter separation scenario discussed below since the rate of the weak processes is much smaller than the rate of strong interactions.}. In the picture of a dilute gas of quarks and hadrons, this would result in a difference between reflection and transmission coefficients for particles carrying baryon and anti-baryon numbers. For example, for a quark incident in the QGP phase the probability ${\cal P}_r$ of reflection back would be different from that ($\bar {\cal P}_r$) of the antiquark. Since the interactions between quarks and hadrons are strong, it is conceivable to assume that the CP(T) asymmetry in reflection coefficients $ \Delta {\cal P} \equiv {\cal P}_r-\bar {\cal P}_r$ is of the order of  
\be
\Delta {\cal P}_0 \simeq \kappa \frac{\mu_Q}{T}\simeq  \frac{\kappa L}{2T^3}\simeq  5\kappa\Delta_L\,,
\label{eq:delp}
\ee
where we used the entropy density at $T=T_c$, $s\simeq \frac{2\pi^2}{45}g^*T^3$ with $g^*\simeq 24$. The asymmetry $\Delta {\cal P}_0$ should be zero if the flavour symmetry were exact, we expect it to behave as $m_s^2/T_c^2$ for $m_s\to 0$.

In a complete thermal equilibrium, with the domain wall at rest, this dynamical feature does not lead to any substantial consequences: the asymmetries in hadronic flavours in different phases can be found by standard methods of equilibrium thermodynamics (see, e.g. \cite{Witten:1984rs} for a sample computation) using the same chemical potentials for conserved charges across the interphase boundary.  Now, if the domain wall is moving with velocity $v=\dot{r}$, the difference in reflection probabilities together with non-equilibrium induced by the bubble wall motion will lead to the baryon current $J$ of the order of
\be
J \sim -  \Delta {\cal P}_0 v n_q~,
\ee
floating inside the droplets of the QGP (given $|\Delta_L|\gg\Delta_B$ the sign of this current is irrelevant for what follows).  Here  
\be
n_q \simeq  g_{\rm QCD}^* n_f=  \frac{3\zeta(3)}{4\pi^2}g_{\rm QCD}^*T^3 \simeq 0.9T^3
\label{eq:nh}
\ee
is an estimate of the total density of the strongly interacting quarks and antiquarks in the QGP (the zero mass free fermionic concentration is $n_f=3\zeta(3)/(4\pi^2)T^3$). So, the equation (\ref{nBeq}) for the baryon number density receives an additional contribution associated with this flux, 
\be
\label{nBCP}
\frac{\partial n_B^d(t)}{\partial t} = -\frac{3\dot r}{r} n_B^d(t) - \dot r n_q  \Delta {\cal P}_0 S/V\,,
\ee
where $V=4/3\pi r^3$ and $S=4\pi r^2$ are the volume and surface of the droplet respectively. The solution  of (\ref{nBCP}) reads
\be
\label{nBCPsol}
n_B^d(t) =  n_B^d(t_0) \left[\frac{r(t_0)}{r(t)}\right]^3 + 
\Delta {\cal P}_0 n_q \left[\frac{r(t_0)^3}{r(t)^3}-1\right]\,.
\ee
For the baryon asymmetry $n_b$ of the bulk exterior to the QGP droplet, corresponding to the hadronic phase, one gets in the full analogy
\be
\label{nbulk}
n_b(t) =  n_B^d(t_0) \frac{V_0}{V_b} -  \Delta {\cal P}_0 n_q \left(\frac{V_0}{V_b}-1\right)\,,
\ee
where $V_b$ is the volume occupied by the hadronic phase, and $V_0$ is its initial value at the moment of the bubble percolation. Once the size of the droplet at some time $t_1$ gets just $25\%$ smaller than its initial value, the value of the baryon asymmetry inside the droplet will increase to the value of the order of initial lepton asymmetry, $n_B^d (t_1)\simeq \Delta {\cal P}_0 n_q$. This large asymmetry may lead to even higher asymmetry in the reflection coefficients, 
\be
\label{kappa1}
\Delta {\cal P} \simeq \kappa_1 \frac{n_B^d}{n_q} \, ,
\ee
where $\kappa_1$ is another unknown parameter, similar to $\kappa$. Using eq. (\ref{kappa1}) in (\ref{nBCP}) one finds that the baryon density inside the QGP droplet increases as
\be
\label{nBCPsolnew}
n_B^d(t) =  n_B^d(t_1) \left[\frac{r(t_1)}{r(t)}\right]^{3+3\kappa_1} \,,
\ee
reaching the nuclear density at $r(t)\simeq  r_0 {\cal P}_0^{\frac{1}{3(1+\kappa_1)}}$.  

The solutions (\ref{nBCPsol},\ref{nBCPsolnew}) correspond to the matter-antimatter separation. The total baryon number of the Universe does not change but gets unequally distributed in the domains (droplets) occupied by the QGP and hadronic matter, the first type carrying an excess of baryons and the second - an excess of antibaryons (or vice versa, depending on the sign of $ \Delta {\cal P}$). Moreover, the baryon asymmetry in the droplets of the QGP can get much larger than the initial lepton asymmetry once the volume fraction occupied by QGP shrinks to $\sim {\cal P}_0^{\frac{1}{(1+\kappa_1)}}$  - an effect we will explore later in Section \ref{sterile}. 

The effect of matter-antimatter separation goes away if the asymmetry in reflection coefficients $ \Delta {\cal P}_0$ is smaller than the average baryon asymmetry of the Universe. In this case, the decrease of the volume of the QGP droplets tends to increase the asymmetries in the individual quark flavours. However, these asymmetries are diluted by the weak reactions of the type $\bar{u}d\to \mu^+\nu_\mu$, $\bar{u}s\to \mu^+\nu_\mu$. As neutrinos can go easily out of the droplets, and the latter reactions are faster than the rate of the droplet shrinking $\dot r/r \sim H$ ($H$ is the Hubble rate), the asymmetries in quark flavours remain at the level of the average lepton asymmetries in the Universe and do not grow with shrinking of the droplets. 

Of course, the assumption that the baryon number cannot leak out of the QGP droplets is presumably too strong and is likely to be wrong at the end of the phase transition, when the density of the baryonic charge approaches that of nuclear density. The unknown strong dynamics prevented us (and the previous workers, e.g. \cite{Witten:1984rs,Kajantie:1986hq}) from making any definite conclusions concerning this point. We would like just to mention that the qualitative remark of \cite{Witten:1984rs} remains in force: the critical temperature of the QCD phase transition with non-zero baryon number decreases when the baryonic chemical potential increases, meaning that the QGP droplets may survive till the temperatures smaller than $T_c$, say $T_c/2$ \cite{Witten:1984rs}.

The mechanism discussed above resembles a lot the scenario of domain wall electroweak baryogenesis (for a review see \cite{Cohen:1993nk} and references therein). In the case of the first-order electroweak phase transition, the intrinsic CP-violation in interactions of quarks, leptons, or other hypothetical fermions with the domain walls also leads to the separation of fermionic number. An excess of fermions over antifermions forms inside the bubbles of a new Higgs phase, whereas the situation is opposite outside the bubbles, in the symmetric phase of the electroweak theory. The fermionic number outside the bubbles is eaten up by equilibrium sphaleron transitions \cite{Kuzmin:1985mm}, which are inactive inside the bubbles if the phase transition is sufficiently strongly of the first order \cite{Shaposhnikov:1986jp,Shaposhnikov:1987tw}. This process leads eventually to the bulk baryon asymmetry after the bubbles with symmetric phase disappear. In the QCD case the CP asymmetry is induced by the electric charge chemical potential and the sphaleron processes are not effective and thus cannot change the baryon number in either phase.

Clearly, the matter-antimatter separation scenario strongly deviates from the standard one in which the universe is homogeneous and isotropic since inflation. An obvious question is about the Big Bang Nucleosynthesis. The sufficiently short-scale inhomogeneities disappear by the nucleosynthesis time \cite{Applegate:1987hm,Jedamzik:1993dc,Jedamzik:1994ux,Kurki-Suonio:1996wfr}  via the combined action of neutrino inflation and neutron diffusion. Baryon number fluctuations affect BBN provided they are sizable enough over the neutron diffusion scale ($3\times 10^5$ cm) at the onset of nucleosynthesis at $T\simeq 100$ keV.  The neutron diffusion scale, blue-shifted to the QCD phase transition scale $T_c \simeq 160$ MeV, becomes, $L_{\rm diff}(T_c) = 2$ m, somewhat larger than the expectation for the distance between the bubble centres, providing an estimate of the scale of baryon number fluctuations. It would be interesting to see whether this may change the BBN predictions and thus put bounds on the possible parameters of the QCD phase transition and lepton asymmetry of the Universe.

Before coming to a possible cosmological consequence of the matter-antimatter separation scenario we will consider in the next Section a much more speculative possibility to have matter-antimatter separation, which would work even without the presence of (large) lepton asymmetry.

\subsection{Omnes phase transition}
\label{omnes}

Back in 1969 Omnes \cite{Omnes:1969ax} proposed an idea of matter-antimatter separation on cosmological scales in an attempt to understand why we have only baryons in the local vicinity of our solar system.  He argued that if baryons and antibaryons are repulsed from each other at small distances, the phase diagram of the strong matter allows the existence of two phases, one with an excess of baryons and another with an excess of antibaryons. If true, this would mean the spontaneous breaking of charge-conjugation symmetry in hadronic matter. According to \cite{Omnes:1972vq,Cisneros:1973ws}, these phases can coexist at some interval of temperatures around $\sim 350$ MeV, and the Universe's evolution may lead to the separation of domains of matter and antimatter at the cosmological scales. 

As was noted by many researchers (see, e.g. \cite{Dolgov:1991fr,DeRujula:1997pz}), the idea fails for many reasons. First, the horizon scale at temperatures $\sim 350$ MeV is too small to create any structures with masses exceeding the solar mass. Second,  theoretical computations of nucleon dynamics at these temperatures cannot be reliable due to the strong coupling. Moreover, the domain of temperatures found by Omnes is already in the QGP region, where quarks and gluons provide a better description of dynamics. 

\begin{figure}[tb]
    \centering
    \includegraphics[width=0.46\textwidth]{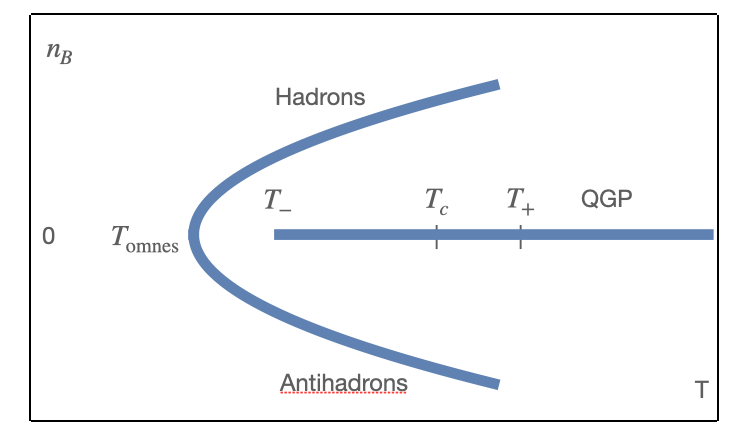}
    \caption{Omnes type QCD phase diagram. In addition to the QGP phase with vanishing baryon number density there are two distinct hadron phases with opposite baryon number densities. The Omnes phase exists at $T_{\rm omnes}<T<T_+$, and the QGP phase at $T>T_-$.}
    \label{fig:omnes}
\end{figure}

Still, it seems to us that the Omnes-type phase diagram such as the one presented in Fig. \ref{fig:omnes}, though extremely exotic,  cannot be completely excluded by the present state of art lattice simulations\footnote{Of course, there are all sorts of other logical possibilities here. For instance, the Omnes phase transition may happen already in the QGP phase, leading to the separation of quarks and antiquarks rather than baryons, with the quark-hadron phase transition occurring later. We will not go through all possibilities as the results are qualitatively the same.}.  So, we find it interesting to discuss the Universe's evolution if it is indeed realised, but we cannot put any argument for why {\em it should be realised}.

As we have strong coupling, it would be natural to assume that at the critical temperature, the baryonic density of the Omnes states is of the order of the nuclear density, i.e. much larger than the baryon and lepton asymmetries of the Universe. In this case, the breaking of the degeneracy between the minima with different baryon numbers will be small and not essential for the initial stages of the QCD phase transition. The Omnes phase transition would start somewhat below $T_c$, as in the discussion in Section  \ref{separation}. However, now two types of bubbles of the hadronic phase will be nucleated. Approximately half of them will carry a positive baryon number, and another half -- negative baryon number. The bubbles will grow and at some moment start to percolate, but now with the simultaneous existence of all three possible phases. The QGP phase will be eaten by the hadronic bubbles and cease to exist at temperature $T_-$ at the latest, and we are left with domains of positive and negative baryonic number densities. The presence of lepton and baryon asymmetries, resulting in the breaking of the degeneracy, will lead to somewhat faster growth of the droplets that have lower free energy. At the temperature $T_{\rm omnes}$ (Fig.  \ref{fig:omnes})  the spontaneous C-breaking comes to an end, and the Universe will be in the hadronic phase, with large inhomogeneities in the distribution of the baryon number, with a typical distance scale of the order of the initial separation between the bubble centres. At this temperature, roughly half of the space will carry a positive baryon number, and another half will be rich in antibaryons, with baryon asymmetry of the order of $\pm 1$ in each of the domains.  This is to be compared with the scenario of Section \ref{LA}, where in the larger part of the space we have an excess of (say) matter with amplitude $\sim  \Delta {\cal P}_0 \ll 1$,  and in a smaller part of the space occupied by the QGP an excess of antimatter. 

Most probably, the BBN can constrain the parameters of the Omnes-type phase transition (such as the average distance between the bubbles and the amplitude of the baryon asymmetry in different C-odd phases). This is not attempted in our paper.  Instead, in the next Section we will consider possible cosmological consequences of the matter-antimatter separation associated with the production of Dark Matter sterile neutrinos.

\section{QCD phase transition and production of sterile neutrino dark matter}
\label{sterile}

\subsection{Matter potential and resonance}

 For the mechanisms discussed in the literature so far, the potential was mainly determined by the lepton asymmetry of the Universe, as the baryon contribution to it is negligibly small. For the matter-antimatter separation scenarios, it is the baryon asymmetry which plays a crucial role. 

The neutrino matter potential is given by \cite{Notzold:1987ik,Ghiglieri:2015jua}  for the electron neutrino
\bea
\label{nupot}
V =\sqrt{2} G_F\left[2\Delta_{\nu_e}+\Delta_{\nu_\mu}+\Delta_{\nu_\tau}
+\left(\frac{1}{2}+2\sin^2\theta_W\right)\Delta_e-\left(\frac{1}{2}-2\sin^2\theta_W\right)\Delta_\mu\right.
\nonumber\\
\left.+\left(\frac{1}{2}-\frac{4}{3}\sin^2\theta_W\right)\Delta_u -
\left(\frac{1}{2}-\frac{2}{3}\sin^2\theta_W\right)\Delta_d-
\left(\frac{1}{2}-\frac{2}{3}\sin^2\theta_W\right)\Delta_s\right]\,,
\eea
where $\Delta_i$ are asymmetries in the concentrations (including spin and colour) of the corresponding fermion flavour, for instance, $\Delta_e\equiv n_e-n_{\bar e}$. For the temperatures of interest $T\sim 200$ MeV, the contribution of $\tau$-lepton can be neglected. Similar expressions can be written for other types of neutrinos and also in the hadronic phase \cite{Notzold:1987ik}. Accounting for the neutrality of the plasma expressed by eq. (\ref{neutr}), one can get from eq. (\ref{nupot})
\be
\label{nupot1}
V=\frac{G_F}{3\sqrt{2}} (7L-3B)\,
\ee
which reduces at $L \ll B$ to
\be
\label{bLbar}
V \approx -\frac{G_F}{\sqrt{2}} B\,.
\ee

To get a better feeling of the numbers involved, we can write the baryon number density as  
\be
\label{norm}
B = \eta_B n_{h,q}, 
\ee
where $\eta_B$ is the baryon asymmetry, and $n_h$ and $n_q$ are the equilibrium number of baryons at zero baryon asymmetry in the hadron and QGP phases respectively.  In the hadronic phase, $n_h$ can be estimated with the use of the so-called non-interacting hadron gas approximation, counting the low-lying baryon and antibaryon states (proton, neutron, $\Lambda$-hyperon and their excitations). At a temperature of the phase transition $T  = 160$ MeV,  it gives
\be
\label{nh}
n_h\simeq 0.03 T^3\,.
\ee
In the QGP phase, $n_q$ is defined in eq. (\ref{eq:nh}),  $n_q \simeq 0.9 T^3$.

Depending on the sign of $V$, there is a level crossing between the sterile neutrino and active neutrino or corresponding antineutrino at a certain momentum (called resonant) $p$, given by
\be
\label{reson}
p_{\rm res}  \approx  \omega_{\rm res} \approx \frac{M_1^2}{2|V|}~.
\ee
The transition probability of neutrinos (or antineutrinos) with this momentum into sterile states is enhanced, leading to the resonant production of $N$. Plugging $V$ into (\ref{reson}) we obtain expression for the resonance energy 
\be
\label{reson2}
\omega_{\rm res} = \frac{M_1^2}{\sqrt{2}G_F B} = 
\frac{M_1^2}{\sqrt{2}G_F \rho_B \eta_B}~.  
\ee

The thermal contribution to the potential \cite{Notzold:1987ik} 
\be
\label{thermal}
V_T=\frac{14\pi\omega}{45}\sin^2\theta_W\frac{G_F^2 T^4}{\alpha_{\rm EM}}(2+ \cos^2\theta_W)\,,
\ee
can be neglected for large asymmetries considered in this work (here $\alpha_{\rm EM}$ is the fine structure constant, and $\theta_W$ is the Weinberg angle). 

Let us introduce the dimensionless resonance parameter as  
\be
\label{reso1}
x_{\rm res}\equiv \frac{\omega_{\rm res}}{T} = \frac{M_1^2}{2|V|T} =
\frac{M_1^2}{\sqrt{2}G_F B T}~ . 
\ee
At $T = 160$ MeV according to (\ref{norm},\ref{nh}) we have in the hadronic phase
\be
\label{reso}
x_{\rm res} = \frac{0.15}{\eta_B}\left(\frac{M_1}{7~{\rm keV}}\right)^2\, ,
\ee
and in the QGP phase, according to (\ref{norm},\ref{eq:nh})
\be
\label{resoqgp}
x_{\rm res} = \frac{0.005}{\eta_B}\left(\frac{M_1}{7~{\rm keV}}\right)^2\, ,
\ee
where the sterile neutrino mass is normalized to $7$ keV having in mind the   $3.5$ keV line reported in several papers \cite{Bulbul:2014sua,Boyarsky:2014jta}. 

\subsection{Resonance oscillations in infinite medium}

The physics picture of $\nu - N$ transition is the following: neutrinos oscillate in between the collisions with particles of the medium. At the collisions, the coherence of the oscillating state is broken and after collisions, the active and sterile components start to oscillate independently from the beginning. Correspondingly, the relevant scales are (i) the oscillation length in medium $l_m$, and (ii) the mean free path: $\lambda_\nu = 1/\Gamma$. 

According to this picture, the rate of production of sterile neutrinos can be written as
\begin{equation}
R_N(\omega)  = \frac{n_F(\omega)}{(2\pi)^3}P (\omega) \Gamma , 
\label{eq:ratesphe}
\end{equation}
where $n_F({\omega})$ is the Fermi distribution function, 
and $P (\omega)$ is the $\nu \rightarrow N$ oscillation probability between two collisions
averaged over the distance between the collisions:
\be
P (\omega) = \sin^2 2\theta_m \langle \sin^2\phi_m \rangle. 
\ee
The mixing angle in matter $\theta_m$  is determined by  
\be
\label{eq:mixangle}
\sin^2 2\theta_m = \frac{\sin^2{2\theta}}{R_{MSW}^2}~, 
\ee
where $R_{MSW}$ is the MSW resonance factor:
\be
R_{MSW}^2(\omega) \equiv \cos^2 2\theta  \left( 1 - \frac{\omega}{\omega_{\rm res}} \right)^2 + \epsilon^2
\label{eq:resfac}
\ee
with 
\be
\epsilon^2 \equiv \sin^2 2\theta +  \left( \frac{2 \omega \Gamma}{M_1^2}\right)^2~.    
\label{eq:epsilon}
\ee
Here the last term corresponds to the broadening of the resonance due to inelastic collisions. It can be obtained by adding $-i \Gamma/2$ to the Hamiltonian of evolution, or more consistently, as the inelastic term to the equation for the density matrix.

The half-oscillation phase $\phi_m$ acquired along the distance $x$ equals
\be
\phi_m = \frac{\pi x}{l_m} = \frac{\pi x R_{MSW}(\omega)}{l_\nu}, 
\label{eq:phase}
\ee
and the oscillation length in the medium is
\be
l_m =  \frac{l_\nu}{R_{MSW}(\omega)} = \frac{4\pi \omega}{M_1^2 R_{MSW}(\omega)}.  
\label{eq:oslength}
\ee
Inserting expression for probability into (\ref{eq:ratesphe}) we obtain the rate of N-production 
\begin{equation}
R_N = \frac{n_F(\omega)}{(2\pi)^3 } \Gamma  \frac{1}{R(\omega)^2}
\sin^2 \phi_m . 
\label{eq:Nrateph2}
\end{equation}
The quantity $\epsilon$  (\ref{eq:epsilon}) can be rewritten as 
\begin{equation}
\epsilon^2 = \sin^2 2\theta \left[\left(\frac{l_R}{2\pi \lambda_\nu}\right)^2 + 1 \right]~,
\label{eq:twoterms}
\end{equation}
where $l_R = l_\nu/2\theta$ is the oscillation length in resonance (in the absence of collisions), $l_R  \gg l_\nu, l_0$. In the epoch of the QCD phase transition $\lambda_\nu \ll l_R/2 \pi$ and the second term in (\ref{eq:twoterms}) can be neglected. Then the expression for the rate of $N-$production (\ref{eq:Nrateph2}) becomes explicitly (taking $c_{2\theta} = 1$)  
\begin{equation}
R_N = \frac{n_F}{(2\pi)^3} 4\theta^2 M_1^4 \frac{\Gamma}
{(M_1^2 - 2\omega V)^2 + (2\omega \Gamma)^2} \sin^2 \phi_m.
\label{eq:Nrateint}
\end{equation}

The key feature of these oscillations is that due to the smallness of vacuum mixing the oscillation length in matter changes with $E$ by many orders of magnitude:  in resonance $l_m = l_R = l_\nu/\sin 2\theta \approx l_\nu/2\theta$,  while below resonance $l_m \approx l_\nu$, and above resonance $l_m \approx l_0$,  $l_R \gg l_\nu$.  So, $l_m \gg \lambda_\nu$ in resonance, while $l_m \ll \lambda_\nu$ outside the resonance. This means that outside the resonance the oscillations are averaged between two collisions, and consequently $\sin^2 \phi_m = 1/2$. The resonance peak exists but its width and height are determined by collisions. The oscillation phase between two collisions equals
\begin{equation}
\phi_m = \pi \frac{\lambda}{l_m}  \approx \frac{\lambda}{2}
\sqrt{\left(\frac{M_1^2}{2\omega} - V \right)^2 +  \Gamma^2 }
\label{eq:oscph}
\end{equation}
and in the resonance $\phi_m \approx  \lambda \Gamma/2  \approx {1/2}$.

Here we computed the rate of production of $N$ by a single active neutrino. If we neglect the term $(M_1^2 \sin 2\theta)^2$ and assume for simplicity the same potential $V$ for all neutrino species (which is justified if the neutral current scattering on quarks dominates), then summation over active neutrinos is reduced to considering $\theta^2$ as the overall vacuum mixing angle squared: 
$$
\theta^2 \equiv \sum \theta_\alpha^2~. 
$$ 

Using relation 
\begin{equation}
\frac{\alpha}{(x_{\rm res} - x)^2 + \alpha^2} = \pi \delta(x - x_{\rm res}) + O(\alpha), 
\label{eq:appr}
\end{equation}
we can rewrite the rate (\ref{eq:Nrateint}) as   
\begin{equation}
R_N \approx \frac{n_F}{(2\pi)^2}
\frac{\theta^2 M_1^4}{2 \omega V} \delta (\omega - \omega_{\rm res}) \langle \sin^2 \phi_m \rangle 
= \frac{n_F }{(2\pi)^2}
\theta^2 M_1^2 \langle \sin^2 \phi_m \rangle  ~\delta(\omega - \omega_{\rm res}),  
\label{eq:Nrateint}
\end{equation}
where we used (\ref{reson}).  For $\sin^2 \phi_m = 1/2$ it coincides with the expression obtained  from the first principle computation of the resonance contribution in \cite{Laine:2008pg} 
\be
\label{ratenus} 
R_N = \frac{n_F(\omega)}{(2\pi)^2 2\omega}\theta^2 M_1^4 \delta(M_1^2-2\omega V)~.
\ee
This result is valid in the narrow width approximation, $\Gamma_{\nu_e} \ll |V|$. 

For very high interaction rate: $\Gamma  \gg    2\pi /l_0 \gg 2\pi/ l_R$, (wide width) the resonance peak essentially disappears and we can not use the $\delta-$ function approximation (\ref{eq:appr}). Formally the oscillation length becomes $l_m \approx 2\pi /\Gamma$ and the oscillation phase acquired between two collisions is $\phi_m = 1/2$. The rate of $N-$production equals according to (\ref{eq:Nrateint})
\begin{equation}
R_N = \frac{n_F}{(2\pi)^3}  \frac{\theta^2 M_1^4 }{\omega^2 \Gamma} \langle \sin^2 \phi_m \rangle 
\label{eq:rn2}
\end{equation}
with  $\sin^2 \phi_m \leq 1/4$. In the opposite extreme case of  very low interaction rate, $\lambda_\nu \gg  l_R/2\pi$, the vacuum term $(M_1^2 s_{2\theta})^2$  in the denominator of (\ref{eq:Nrateph2}) dominates and we obtain
\begin{equation}
R_N = \frac{n_F}{(2\pi)^2}  \frac{\Gamma \theta^2 M_1^2 }{4V} \delta (\omega - \omega_{\rm res}) =
\frac{n_F}{(2\pi)^2}  \frac{\Gamma \theta^2 \omega_{\rm res} }{2} \delta (\omega - \omega_{\rm res}).
\label{eq:rn}
\end{equation}
Thus,  the rate of $N$ production has the following dependence on $\Gamma$: for low $\Gamma$, $R_N \propto \Gamma$, in the intermediate range the rate does not depend on $\Gamma$ and for large $\Gamma$, $R_N \propto 1/\Gamma$.

To get the total number density of sterile neutrinos, $n_N$, the rate $R_N$ in eq. (\ref{eq:Nrateint})
should be integrated over time and momentum:
\be
\label{nusdens}
n_N=   \int dt \int d^3p R_N (\omega) = 4\pi \int dt \int d\omega ~\omega^2 R_N (\omega).
\ee
Integration  over energy  is trivial due to the $\delta-$ function, giving
\begin{equation}
n_N = 
\frac{1}{2\pi} \theta^2 M_1^2 \int dt  n_F(\omega_{\rm res})~
 \omega_{\rm res}^2 \langle 2 \sin^2 \phi_m \rangle.
\label{eq:Nrateint2}
\end{equation}
For $\langle 2 \sin^2 \phi_m\rangle = 1$ the concentration of resonantly produced sterile neutrinos at the time $t$ is given by 
\be
n_N(t) =  \frac{1}{2\pi} \theta^2 M_1^2 T^2
\int_0^{t} dt ~ n_F(x_{\rm res})~ x_{\rm res}^2\,,
\label{eq:Nrateint3}
\end{equation}
which is the same as eq. (2.7) of \cite{Laine:2008pg}, rewritten in terms of time integral rather than the temperature one. 

Assuming that the production of the sterile neutrinos stops at temperature $T$ corresponding to $t$, their concentration at the present epoch can be found with the use of the entropy conservation by a standard computation, giving
\be
n_N(now)=n_N(t) \frac{s(now)}{s(T)}\,,
\ee
where $s(now)/s(T)=g^*(now)/g^*(T)(T_\gamma/T)^3$ with $g^*(now)=3.9$, and $T_\gamma=2.73~K$ being the temperature of the CMB. Thus the ratio of energy density in sterile neutrinos to DM energy density is given by
\be
\label{forfig}
\frac{\Omega_N}{\Omega_{\rm DM}}=\frac{n_N M_1}{\Omega_{\rm DM}}\frac{s(now)}{s(T)}\,.
\ee

For a future comparison,  we present below an estimate of the sterile neutrino DM abundance in the homogeneous situation in the presence of both baryon and lepton asymmetries, see eq. (\ref{nupot1}). In the assumption of standard cosmology without the QCD phase transition and neglecting the temperature dependence of $g^*$ the integral in (\ref{eq:Nrateint3}) can be computed analytically, giving the abundance of the resonantly produced sterile neutrino,
\be
\frac{\Omega_N}{\Omega_{\rm DM}} \simeq 10^{16}\theta^2\left[\frac{1}{g^*}|\Delta_L-\frac{3}{7}\Delta_B|\left(\frac{M_1}{7~{\rm keV}}\right)^2\right]^{3/4}\,
\label{eq:SF}
\ee
and the temperature of the maximal production rate equals
\be
T_{\rm prod}\simeq 380~ {\rm MeV}\left(\frac{M_1}{7~{\rm keV}}\right)^{\frac{1}{2}}\left[\frac{1}{g^*(T)}\frac{6.6\times 10^{-5}}{\Delta_L }\right]^{\frac{1}{4}}\,,
\label{tprod}
\ee
where we used the observed DM abundance $\Omega_{\rm DM} = 1.26 \cdot 10^{-6}$ GeV/cm$^{-3}$.  This formula overestimates to number of sterile neutrinos by a factor of a few. It does not take into account the thermal contribution (\ref{thermal}) to the matter potential, which cuts the resonance at high temperatures. Also, the delta-function approximation to the rate (\ref{eq:rn}) becomes invalid at temperatures large enough, where the resonance is suppressed by the active neutrino collisions.  Finally, it does not incorporate the depletion of lepton asymmetry because of the resonance transitions of active to sterile neutrinos and temperature dependence of $g^*$. 

The most elaborated analysis of the sterile neutrino production in lepton asymmetric homogeneous Universe, accounting for the finite width of the neutrino, the back reaction of produced sterile neutrinos, and non-resonant production can be found in \cite{Ghiglieri:2015jua}.

With these considerations, we are now in a position to estimate the sterile neutrino production in matter-antimatter separating QCD phase transition. We will discuss first a simpler, but more exotic possibility of the Omnes phase transition (Section \ref{omnes}). Then we will turn to a more realistic scenario of Section \ref{LA}.

\subsection{Sterile neutrino DM production at the Omnes phase transition}\label{sterileomnes}

Suppose that the Omnes-type QCD phase transition took place. As we discussed in Section \ref{omnes}, at temperatures below $T_-$, half of the universe is occupied by hadronic matter with the baryon excess, and another part contains an excess of antimatter. In both parts, the resonant production of sterile neutrinos takes place by neutrinos and antineutrinos. Contrary to the homogeneous situations, the sterile neutrinos will be produced in left- or right-handed polarisations, depending on the sign of the baryon asymmetry in a given domain.\footnote{Anyway, the polarization of sterile neutrinos is ``forgotten'' in the course of further evolution of the Universe due to their interaction with the gravitational field of galaxies and clusters of galaxies \cite{Ruchayskiy:2022eog}.} So,  production here is essentially the same as in a homogeneous background.

If characteristics of medium $n$, $T$ do not change substantially during phase transition, then according to (\ref{eq:Nrateint3}) the total number density of $N$ produced during the time $t_{\rm PT}$ is
\begin{equation}
n_N \approx
\frac{n_F(x_{\rm res})}{2\pi}\theta^2 M_1^2 x_{\rm res}^2 T_{\rm PT}^2 t_{\rm PT}~.
\label{eq:Nrateint2}
\end{equation}
We expect that this equation gives a fairly good account of the sterile neutrino production, but in reality, one would need to take into account the change of parameters of the medium during the phase transition (and even after the matter-antimatter domain structure exists). This is not attempted here because of the many uncertainties involved. It is clear that only an order of magnitude estimate can be made at the present stage, as even the mere existence of the Omnes-type phase transition is speculation. We note, however, that contrary to eq.  (\ref{eq:SF}) the depletion of the lepton asymmetry due to resonant transition can indeed be neglected since the matter potential is due to baryon rather than lepton asymmetry. Also, the temperature at which the process of conversion happens is relatively small, justifying the delta-function approximation of the rate and dropping off the thermal contribution to the potential.

Eq. (\ref{eq:Nrateint2}) leads to the sterile neutrino DM abundance
\be
\label{sabun}
\frac{\Omega_N}{\Omega_{\rm DM}} \simeq 4\times 10^{12}  \theta^2 \left(\frac{M_1}{7~{\rm keV}}\right)^3
x_{\rm res}^2 n_F(x_{\rm res}) \frac{t_{\rm PT}}{t_H}\,.
\ee
This formula shows that the matter-antimatter separating Omnes phase transition can easily accommodate 100\% sterile neutrino DM abundance for different choices of mixing angles and masses. For example, keeping in mind the possible detection of $7$ keV sterile neutrino decays by X-ray satellites \cite{Bulbul:2014sua,Boyarsky:2014jta} with $\theta^2 \simeq (0.8 - 5) \cdot 10^{-11}$, the choice  $\theta=1.5\times10^{-11}$, $t_{\rm PT}\simeq  t_H/4$,  and $\eta_B \simeq 1/2$ does the job. An interesting feature of the sterile DM distribution is that it can be much cooler than that of the typical active neutrino with $\langle \omega \rangle \simeq 3.15 T$. Indeed,  according to (\ref{reso}) $\omega_{\rm res}/ \langle \omega \rangle  = x_{\rm res}/3.14  \simeq 0.1$  for $\eta_B \simeq 1/2$, making it essentially the cold dark matter particle, which satisfies easily all the Lyman -$\alpha$ \cite{Boyarsky:2018tvu} and the strongest structure formation constraints of \cite{DES:2020fxi,Enzi:2020ieg,Zelko:2022tgf,Lovell:2023olv}. 

\subsection{Sterile neutrino production in non-uniform medium with  large QGP  droplets}
\label{sterileLA}

Let us assume now that the QCD phase transition goes as described in Section \ref{LA} via the formation of QGP droplets with enhanced baryon number. Also, we take that the average value of the leptonic asymmetry is much larger than the baryon asymmetry but smaller than the critical value $L_{\rm crit}\simeq 6.6\times10^{-5}$ so that no $7$ keV sterile neutrino with the mixing angle $\theta^2=5\times10^{-11}$ can accommodate for all DM in the Universe would be possible if the QCD phase transition is absent.

The key feature of this scenario is that the droplets of QGP  shrink, and consequently,  the baryon number density in them increases. That is,  soon after the time of formation of droplets $t_0$ (percolation time of hadron bubbles), the baryon number density inside the droplets, $n_B^d$, grows like in eq. (\ref{nBCPsol}),  with omitted first term.  The last term in the bracket gives
\be
\label{nbgr}
n_B^d(t) = n_B^d(t_0) \left[\frac{r(t_0)}{r(t)}\right]^3,
~~~~~n_B^d(t_0) = \Delta {\cal P}_0 n_q ~.
\ee
For later times, the baryon density would follow (\ref{nBCPsolnew}).

Depending on the initial size of the QGP droplets at the percolation $t_0$, computations of the sterile neutrino yield proceed in two different ways.

If the initial droplet sizes $r_d$ are much larger the neutrino mean free path $\lambda_\nu$ (say $r_0 \sim$ 1 m while $\lambda_\nu \sim 0.4$ cm (see Section  \ref{LA}), the processes of sterile neutrino production described in the previous subsection take place inside the droplet. We will assume that the baryon number density inside the droplets does not depend on distance and surface effects are negligible. Also, we will neglect oscillations between the droplets, since the baryon and lepton asymmetries there are assumed to be below the critical value. Thus, these domains cannot produce enough sterile neutrino DM.  

To get the number of sterile neutrinos produced inside the droplets, one can repeat the procedure leading to (\ref{eq:Nrateint3}), but accounting for the fact that only the part of space occupied by the droplets, $V_{\rm QGP}$, leads to $N-$ creation. Furthermore, this part decreases with time. The total volume can be presented as $V_{\rm tot} = V_{\rm QGP} + V_{\rm h} \approx 2V_{\rm QGP}(t_0)$, that is, in the initial moment about half of the space is occupied by droplets. Then the QGP part as a function of time equals
\be
\frac{V_{\rm QGP}(t)}{V_{\rm tot}} =
\frac{V_{\rm QGP}(t)}{2V_{\rm QGP}(t_0)}\equiv \frac{1}{2}F(t)\,,
\label{eq:frac}
\ee
and consequently, according to (\ref{eq:Nrateint3})
\be
\label{nNintdr}
n_N(t) =  \frac{\theta^2 M_1^2T^2}{4\pi} \int_{t_0}^t dt \, x_{\rm res}^2 \,
n_F(x_{\rm res})F(t) \, .
\ee

Now the resonance energy depends on the baryon number density in the droplets and therefore on time due to droplet contraction. Indeed,  we have
\be
x_{\rm res}(t) = x_{\rm res}(t_1) \frac{n_B^d(t_1)}{n_B^d(t)}=  x_{\rm res}(t_1) F(t)^{1+\kappa_1}~,
\label{eq:xresd}
\ee
where the initial value of the resonance parameter can be taken at time $t_1$ defined in (\ref{nBCPsolnew}) and $\kappa_1$ in (\ref{kappa1}), and is equal to 
\be
x_{\rm res}(t_1) =  \frac{M_1^2}{\sqrt{2}G_F n_B^d(t_1) T} = 
\frac{M_1^2}{\sqrt{2}G_F n_q \Delta{\cal P}_0T}~.
\label{eq:xresd1}
\ee
where the total density of hadrons $n_q$ is defined in (\ref{eq:nh})  while the difference of reflection coefficients  $\Delta{\cal P}_0$  -- in (\ref{eq:delp}, \ref{kappa1}). 

Thus, the number of created sterile neutrinos depends in an essential way on the evolution of the fraction of the QGP, which is largely unknown. The ratio $\frac{\Omega_N}{\Omega_{\rm DM}}$ is given by (\ref{forfig}) where $n_N$ is to be taken from (\ref{nNintdr}) and $x_{\rm res}(t_1) $ from (\ref{eq:xresd1}).

\begin{figure}[h]
   \centering
    \includegraphics[width=0.46\textwidth]{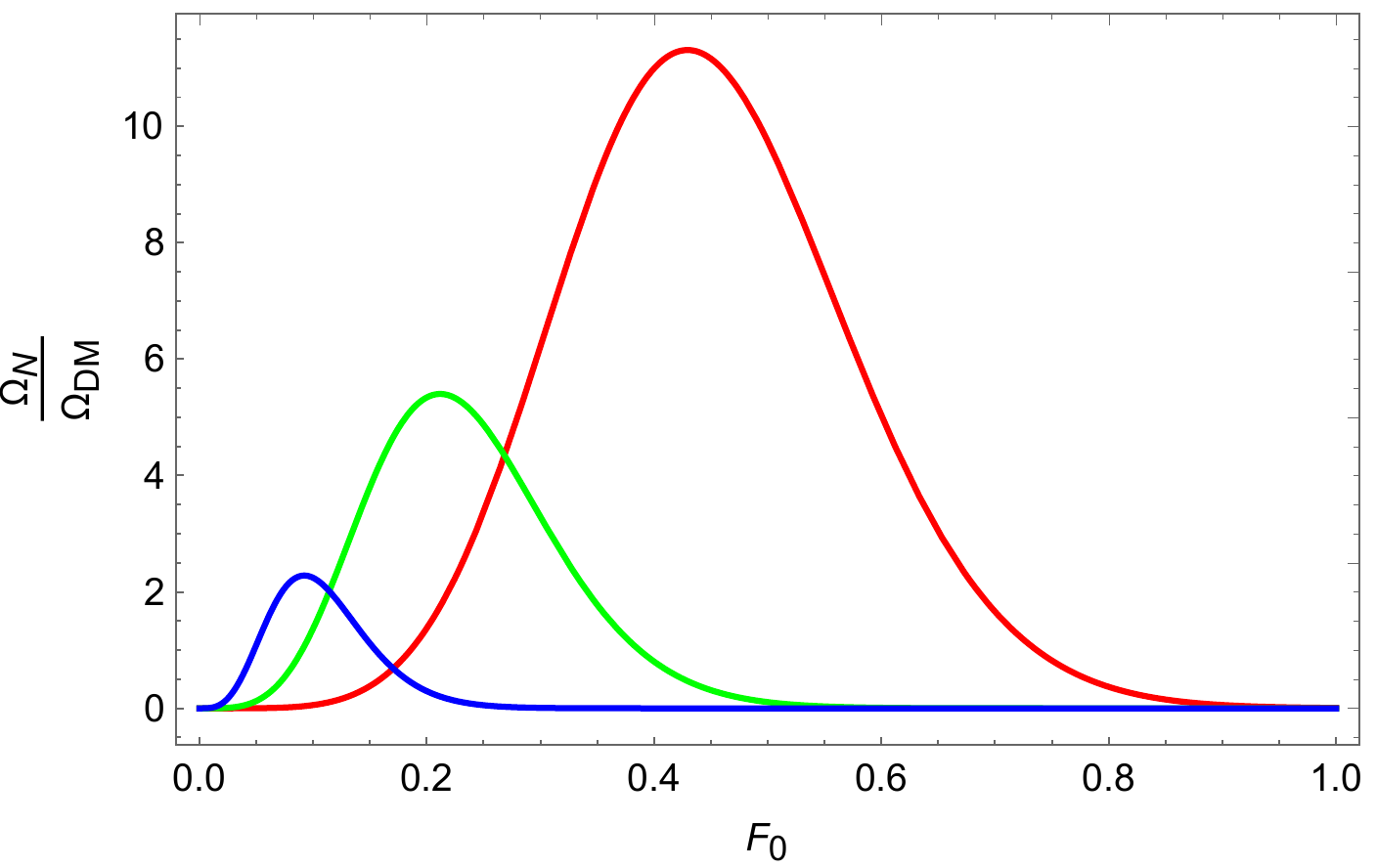}
    \caption{Dependence of the ratio $\frac{\Omega_N}{\Omega_{\rm DM}}$ (vertical axis)  on $F_0$ (horizontal axis) for $M_1=7$ keV, $\theta^2 = 5\times10^{-11}$, and $\Delta_L=L_{\rm crit}$. Red, green and blue curves correspond to $\kappa=\kappa_1=1,~1/2$ and $1/4$ respectively.}
    \label{fig:droplet1}
\end{figure}

A very rough estimate can be derived with a sample function $F(t)$ which starts at $1$ at $t=t_1$, quickly decreases to a constant $F_0$, and then at the end of the phase transition falls to $F(t)=0$, imitating the initial growth of the baryon density, then slowing down the droplet shrinking because of the extra pressure acting from inside, and finally the droplet disappearance. The dependence of the sterile neutrino DM abundance defined in (\ref{forfig}) on $F_0$ is shown in Fig. \ref{fig:droplet1}, demonstrating that it can exceed the required DM abundance for $\Delta_L=L_{\rm crit}$ for a wide range of parameters characterising the unknown dynamics of the droplet shrinking. In other words, the QCD phase transition may enhance the sterile neutrino production and may lead to $100\%$ of its abundance even if the lepton asymmetry is smaller than the critical one, though to see if this indeed happens would require the precise knowledge of the droplet dynamics. 

\begin{figure}[h]
   \centering
    \includegraphics[width=0.46\textwidth]{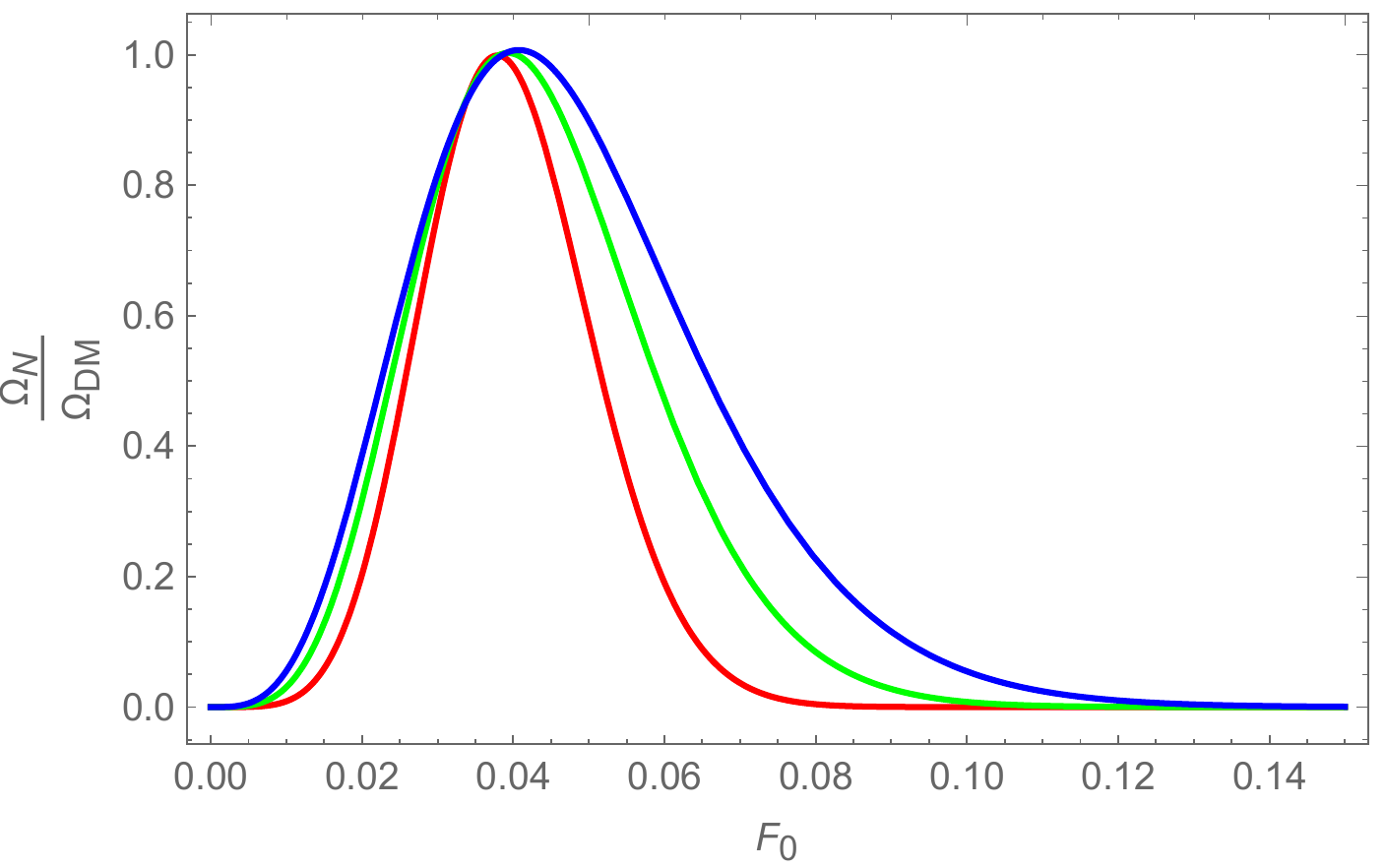}
    \caption{Dependence of the ratio $\frac{\Omega_N}{\Omega_{\rm DM}}$ (vertical axis)  on $F_0$ (horizontal axis) for $M_1=7$ keV, $\theta^2 = 5\times10^{-11}$ for different values of  the ratio $\Delta_L/L_{\rm crit}$ and $\kappa,~\kappa_1$.   Red, green and blue curves correspond to $\kappa=\kappa_1=1; ~\Delta_L/L_{\rm crit}=0.008$, $\kappa=\kappa_1=1/2; ~\Delta_L/L_{\rm crit}=0.08$, and $\kappa=\kappa_1=1/4; ~\Delta_L/L_{\rm crit}=0.4$  respectively.}
    \label{fig:droplet2}
\end{figure}

To get a feeling of how small the required asymmetry could be, we present in Fig. \ref{fig:droplet2} the dependence of the sterile neutrino abundance on $F_0$ for different values of the ratio $\Delta_L/L_{\rm crit}$, which are chosen in such a way that the maxima of the corresponding curves reach one at some value of $F_0$. Even the asymmetries smaller than the critical one by a factor of $100$ are possible for the specific choice of $F_0\simeq 0.04$. 

A somewhat more involved time dependence of the QFP fraction can be written as
\be
\label{fracx}
F(t)=\left(\frac{t_f-t}{t_{\rm PT}}\right)^\alpha\, ,
\ee
where $t$ is time, $t_f$ ($t_i$) corresponds to the end (beginning) of the phase transition so that $t_{\rm PT} \equiv t_f-t_i$. We leave the power $\alpha$ to vary between $1$ and $3$: $\alpha=1$ corresponds to the behaviour found in \cite{Kajantie:1986hq} by solving FRW equations during the QCD phase transition assuming the entropy conservation, whereas  $\alpha=3$ would correspond to the motion of the bubble walls with the constant velocity, accounting to friction, often found in electroweak baryogenesis (see, e.g. \cite{Khlebnikov:1992bx}). It is difficult to argue that the QGP fraction is indeed described by (\ref{fracx}) because of poor knowledge of the dynamics of the phase transition.

\begin{figure}[h]
   \centering
    \includegraphics[width=0.46\textwidth]{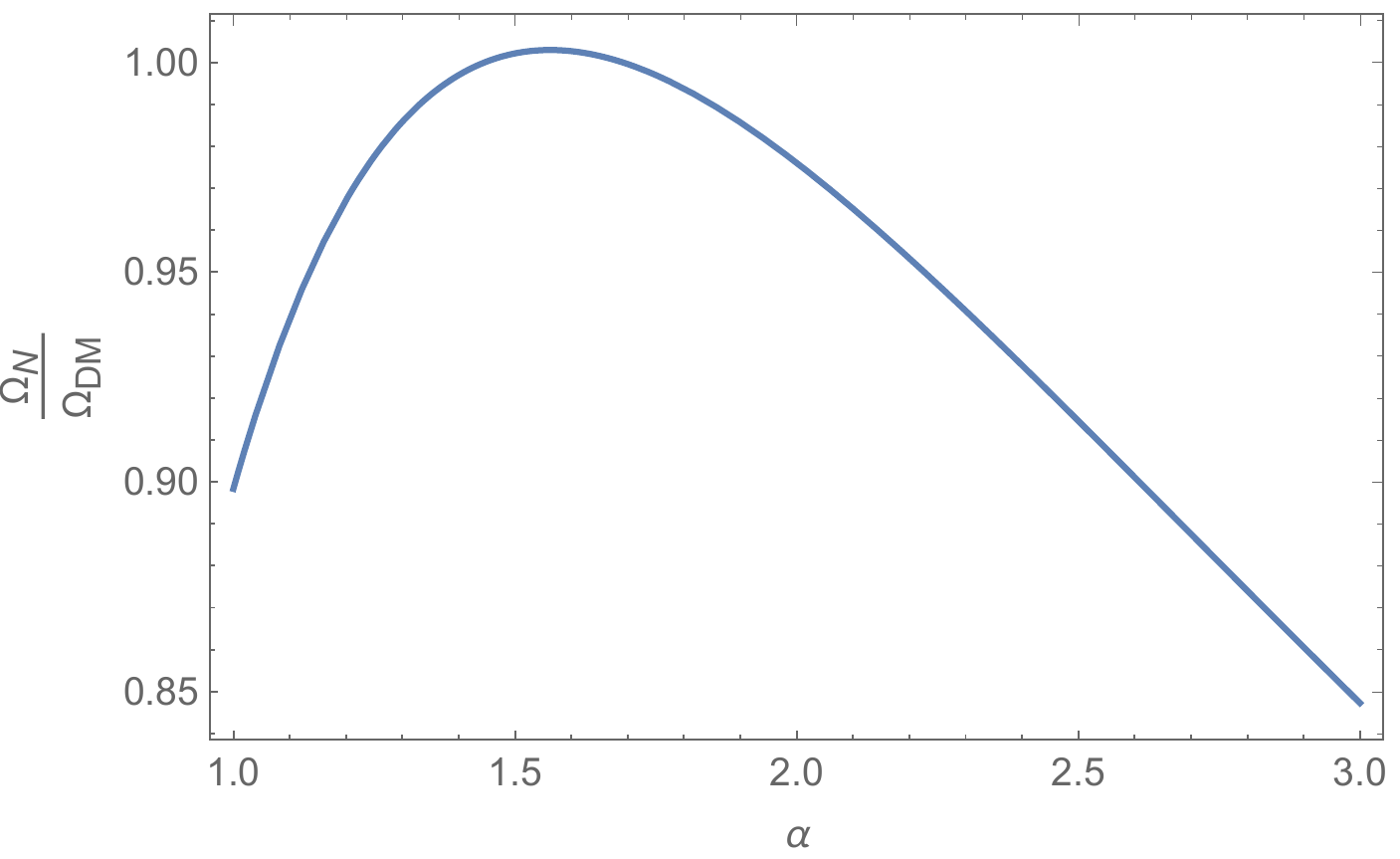}
    \caption{Dependence of the ratio $\frac{\Omega_N}{\Omega_{\rm DM}}$ (vertical axis)  on $\alpha$ (horizontal axis) for $M_1=7$ keV, $\theta^2 = 5\times10^{-11}$, and $\Delta_L=L_{\rm crit}/4$.}
    \label{fig:droplet3}
\end{figure}

In Fig. \ref{fig:droplet3} we show the sterile neutrino DM abundance as a function of $\alpha$ for $\Delta_L=L_{\rm crit}/4$, demonstrating that certain droplet dynamics can generate enough sterile neutrinos for asymmetries smaller than the critical one, in this case for $\alpha\simeq 1.5$.

\subsection{Oscillations in medium with small droplets}
\label{smalldrop}

Let us consider now the situation in which the initial size of the droplets is smaller than the neutrino mean free path $\lambda_\nu$. Since $\lambda_\nu \propto 1/n_{tot}$   and inside the droplets total number density $n_{tot}$ may only weakly increase the inequality $r_d < \lambda_\nu$ will maintain during PT. In this case, a given active neutrino can cross two or more droplets.

Now we have the following relation between the scales:
\be
l_\nu \sim l_0 \ll r_d \ll  l_\nu^R,
\label{eq:lrel}
\ee
where $r_d$ is the radius of the droplet. If the droplets have a spherical form the average distance crossed by neutrinos is $\bar{r} = (\pi/2) r_d$. At the conditions (\ref{eq:lrel}) the oscillation phase in the MSW  resonance: $\phi_R(\bar{r}) \ll 1$, while in the tails above and below resonance energy $\phi \gg 1$. Therefore the  probability of the $\nu \to N$ conversion in the energy range close to the resonance energy $E\simeq E_{\rm res}$ is given by
\be
\label{resen}
P \approx \sin^2 \theta \left(\frac{\pi \bar{r}}{l_\nu}\right)^2,
\ee
$R$ cancels and the probability does not depend on the matter density (the so-called "vacuum mimicking" regime  \cite{Akhmedov:2000cs,Yasuda:2001va}. Exactly at  $\omega = \omega_{\rm res}$
\be
\label{conv}
P_{\rm res} \approx 4 \pi^2 \theta^2 \left[\frac{\bar{r}}{l_\nu (\omega_{\rm res})} 
\right]^2\, .
\ee

The probability as a function of energy has a peak inscribed into the MSW  resonance peak with maximal value $\approx P_{\rm res}$  (\ref{conv}). The width of the peak, $\delta \omega = \omega_W - \omega_{\rm res}$,  is determined approximately by the condition that the oscillation phase in matter equals $\pi/2$:
\be
\phi_m (\omega_W) = \frac{\bar{r}  M_1^2 R(\omega_W)}{4\omega_W} 
= \frac{\pi}{2}. 
\label{eq:widthp}
\ee
The solution of (\ref{eq:widthp}) is
\be
\label{width}
\frac{\delta \omega}{\omega} \equiv
\left(1 - \frac{\omega_W}{\omega_{\rm res}}\right) \approx
\pm \frac{l_\nu(\omega_{\rm res})}{\bar{r}}
\approx    \left(\frac{l_\nu}{2\bar{r}}\right) .    \,
\ee
At $\omega _W$ the peak touches the MSW resonance line. The probabily is zero at $\phi_m = \pi$, {\it i.e.} at $\delta \omega/\omega = {l_\nu}/{r_d}$. At the points  $\omega = \omega_W$, the probability equals $P_W = 16 \theta^2 (\bar{r}/l_\nu)^2$. Consequently,  $P_{\rm res}  / P_W = \pi^2/4 = 2.46$, that is,  $\delta \omega/\omega$ is the width at approximately half of the height of the peak. Notice that in contrast to the case of big droplets here the biggest energy scale $1/\bar{r}$ plays the role of $\Gamma$ and determines the properties of the peak.  

The $\nu \rightarrow N$ oscillation probability averaged over the spectrum of active neutrinos equals 
\be
\langle P_N \rangle = \frac{1}{n_\nu} 4\pi
\int d\omega \omega^2 \frac{n_F(\omega)}{(2\pi)^3} P_N(\omega)
\approx P_{\rm res} z(2\delta \omega),
\label{eq:pav}
\ee
where $n_\nu = 3\zeta(3)/(4\pi^2)T^3$ is the neutrino concentration (one chiral degree of freedom) and
\be
\label{fraction}
z(2\delta \omega) = \frac{1}{n_\nu} 4\pi \int_{2 \delta \omega}  d\omega \omega^2 
n_F(\omega)  =
\frac{4}{3\zeta(3)}\left(\frac{\omega_{\rm res}}{T}\right)^3
n_F(\omega)\frac{\delta \omega}{\omega}\, 
\ee
is the fraction of active neutrinos in the energy interval  $2\delta \omega$. Here we have taken into account the contribution of peak of $P(\omega)$ to the integral. The contribution of the non-resonance tails can be estimated using the averaged oscillation probability in a vacuum and $\delta \omega /\omega \sim 1$. This gives $\langle P_N \rangle \sim 2 \theta^2$, smaller than resonance contribution by a factor $2\pi^2 \bar{r}/l_\nu \gg 1$ and can be neglected. Inserting the fraction (\ref{fraction}) into (\ref{eq:pav}) we obtain
\be
\label{ptot}
\langle P_{N} \rangle \approx 
\frac{2\pi}{3\zeta(3)}\frac{\theta^2 M_1^2 
\bar{r}}{T} \left(\frac{\omega_{\rm res}}{T}\right)^2 n_F(\omega_{\rm res})\, .
\ee
This is averaged over the energy probability that the active neutrino produces a sterile neutrino while crossing a droplet.

Using $\langle P_{N} \rangle$ the total production rate of sterile neutrino number density $R_N$ can be performed in two different ways.

1. Consider a given active neutrino propagating in a medium with QGP droplets with number density $n_{drop}$. The rate of $N-$ production equals
\be
R_N = \langle P_{N} \rangle P_{drop} n_\nu,
\label{eq:nNa}
\ee
where $P_{drop}$ is the probability that active neutrino hits the droplet in the unit of time:
\be
P_{drop} =  \sigma_{drop} n_{drop}.
\label{eq:nNa}
\ee
Here $\sigma_{drop} = \pi r_d^2(t)$ is the area (cross section) of the droplet. Inserting all the factors into (\ref{eq:nNa}) gives
\be
R_N = \langle P_{N} \rangle  \pi r_d^2(t) n_{drop} n_\nu~.
\label{eq:nNa1}
\ee
This computation is similar to the computation of the process of scattering $\nu+{\rm droplet}\to N+{\rm droplet}$.

2. Consider a given droplet and compute the number of sterile neutrinos produced in this droplet in the unit time: 
\be
n_N = \langle P_{N} \rangle F_\nu n_{drop},
\label{eq:nNb}
\ee
where $F_\nu$ is the active neutrino flux entering the droplet:   $F_\nu = 0.5 n_\nu S_{drop} =  2\pi r_d^2(t) n_\nu$. Here $S_{drop}$ is the surface of a droplet and factor 0.5 accounts that only half of neutrinos in a given point of the surface enter a droplet. Collecting all the factors we obtain from (\ref{eq:nNb})
\be
n_N = \langle P_{N} \rangle 2\pi r_d^2(t) n_\nu  n_{drop} \,,
\label{eq:nNb1}
\ee
which coincides  with expression (\ref{eq:nNa1}). 

The number density of droplets $n_{drop} = 1/d^3$. Here $d$ is the distance between the centres of droplets. It does not change with time,  while the radius of the droplet decreases with time $d \approx 2 r_d(t_0)$ and $r_d(t_0)$ is the initial radius of the droplet.  Inserting expression for $n_{drop}$ into (\ref{eq:nNa1}) and integrating over time  we obtain
\be
n_N = \pi {n_\nu} \int dt
\langle P_{N} \rangle r_d^2(t) \frac{1}{(2 r_d(t_0))^3}.
\label{eq:nNbtime}
\ee
Then using explicit expression for $\langle P_{N} \rangle$ from  (\ref{ptot}) we get
\be
n_N  = \frac{\theta^2 M_1^2}{16}  \int dt \omega^2 n_F(\omega)
\left(\frac{r_d(t)}{r_d(t_0)}\right)^3.
\label{eq:nNfinal}
\ee
This result coincides up to a factor of the order one with expression for $n_N$ in the case of large droplets (\ref {nNintdr})  (16 in the denominator should be substituted by $4 \pi$),  which is not accidentals. Consequently, further integration over $t$ proceeds in the same way as in the section \ref{sterileLA}. Therefore, the oscillations of active neutrinos in the small droplets of QGP with large baryon density may also lead to enhanced sterile neutrino production.

As previously,  the required lepton asymmetry to accommodate the totality of DM may be considerably smaller (say, a factor of 10 or even 100, as we saw in the previous section) than the critical one, though a reliable estimate of the lower bound is hardly possible with the current uncertainties.  In comparison with the resonant production in the homogeneous universe  (\ref{eq:SF}), the gain is explained by the increased neutrino potential in the droplets of the QGP at the QCD phase transition, in addition, the parametric dependence of the abundance on the mass of the sterile neutrino and asymmetry in both cases are different from each other.

\section{Discussion and outlook}
\label{discuss}

In this paper, we considered the production of DM sterile neutrinos in an inhomogeneous Universe, temporarily filled by matter-antimatter domains at temperatures of the order of the QCD scale. 

First, we argued that rejecting the possibility of the first-order QCD phase transition might be premature. The conclusion on the absence of the phase transition is based on lattice simulations performed up to now, and the use of much larger lattices may be needed to clarify its nature. Our point of view goes against the consensus reached in this domain of high-energy physics. Though we can put no arguments saying that the QCD phase transition should take place, we think that it makes sense to assume that it happened, and explore its possible consequences. 

Second, we argued that the QCD phase transition may lead to matter-antimatter separation. This is a speculation about the unknown strong dynamics: neither the existence of the Omnes-type phase transition nor the CP-breaking properties of the interphase tension between the hadronic phase and the QGP in the presence of lepton asymmetries were derived from the first principles and are simply assumed.  At the same time, we found no arguments which forbid these possibilities. A much better understanding of QCD at temperatures $T\sim 200$ MeV is needed to address these questions.

The most robust part of our work is associated with sterile neutrino DM production in the non-homogeneous situation. If the matter-antimatter separation takes place by these or some other mechanisms (for instance, due to the possible existence of stochastic hypermagnetic fields \cite{Giovannini:1997gp}), the production of sterile neutrino dark matter may be enhanced in comparison with the homogeneous situation. For the Omnes-type phase transition, we found that one can easily accommodate all DM in sterile neutrinos. In another scenario (with two co-existing phases but a CP-violating boundary due to the presence of lepton asymmetry) the DM sterile neutrinos may be effectively produced if the lepton asymmetry exceeds by $\sim 4-5$ orders of magnitude the average baryon asymmetry in the Universe  $\Delta_L\gtrsim 10^{-6}$.  

The latter result may shed light on the mass scale of the heavier HNL's in the $\nu$MSM. A detailed study of \cite{Eijima:2020shs} has shown that the lepton asymmetry of the Universe generated at the freeze in of $N_{2,3}$\footnote{The production of large lepton asymmetries at this moment does not require a delicate fine-tuning between the Majorana mass splitting and the Higgs induced mass splitting \cite{Roy:2010xq}.}  can only reach the necessary value provided the masses of  $N_{2,3}$ are sufficiently small. Indeed, Figures 8 and 9 of  \cite{Eijima:2020shs} show that the asymmetry, greater than, say,  $\Delta_L\gtrsim 10^{-5}$ can be achieved in the $\nu$MSM only for HNL masses between $1$ and $2.5$ GeV if the neutrino mass ordering is normal.  So large asymmetry cannot be generated if the neutrino mass hierarchy is inverted. This may serve as the very first indication of the scale of the mass of HNLs:  the explanation of neutrino masses and baryogenesis, together with BBN constraints only provides a lower bound on their mass, $M_N>140$ MeV (for a review see \cite{Boyarsky:2009ix}). The search for HNLs in this mass domain is potentially possible at the intensity frontier of particle physics, in the experiments like SHiP at CERN SPS \cite{Alekhin:2015byh}.

\section*{Acknowledgments} 

The work of MS was supported by the Generalitat Valenciana grant PROMETEO/2021/083. MS also thanks Harvey Meyer for the discussion of the lattice simulations of the QCD phase transition.

\bibliographystyle{utphys}
\bibliography{Refs}

\end{document}